\documentclass[fleqn,usenatbib]{mnras}
\usepackage{newtxtext,newtxmath}
\usepackage{newtxtext,newtxmath}

\usepackage[T1]{fontenc}
\usepackage{ae,aecompl}

\usepackage{graphicx}	
\usepackage{amsmath}	
\usepackage{hyperref}
\usepackage{natbib}
\usepackage[caption=false]{subfig}
\usepackage{ragged2e}
\usepackage{placeins}
\usepackage{bigints} 
\usepackage{physics}
\usepackage{xcolor}
\usepackage{cleveref}
\usepackage[export]{adjustbox}
\usepackage{ulem}

\newcommand{\figsize}{0.48}

\title[Role of temperature and metallicity in the evolution of thermal instability]{Shatter or not: role of temperature and metallicity in the evolution of thermal instability}
\author[H. K. Das, P. P. Choudhury, P. Sharma]{
Hitesh Kishore Das$^{1}$\thanks{E-mail: hiteshk@iisc.ac.in},
Prakriti Pal Choudhury$^{1,2}$\thanks{E-mail: pp512@cam.ac.uk},
Prateek Sharma $^{3}$\thanks{E-mail: prateek@iisc.ac.in}
\\
$^1$ Department of Physics, Indian Institute of Science, Bangalore 560012, Karnataka, India \\
$^2$ Institute of Astronomy, University of Cambridge, Madingley Rd, Cambridge CB3 0HA, United Kingdom \\
$^3$ Department of Physics and Joint Astronomy Program, Indian Institute of Science, Bangalore, India 560012}

\pubyear{2019}

\begin{document}
\label{firstpage}
\pagerange{\pageref{firstpage}--\pageref{lastpage}}
\maketitle
\begin{abstract}
 We test how metallicity variation (a background gradient and fluctuations) affects the physics of local thermal instability using analytical calculations and idealized, high-resolution 1D hydrodynamic simulations. Although the cooling function ($\Lambda[T,Z]$) and the cooling time ($t_{\rm cool}$) depend on gas temperature and metallicity, we find that the growth rate of thermal instability is explicitly dependent only on the derivative of the cooling function relative to temperature ($\partial \ln \Lambda/\partial \ln T$) and not on the metallicity derivative ($\partial \ln \Lambda/ \partial \ln Z$). For most of $10^4~{\rm K} \lesssim T \lesssim 10^7~{\rm K}$, both the isobaric and isochoric modes (occurring at scales smaller and larger than the sonic length covered in a cooling time [$c_s t_{\rm cool}$], respectively) grow linearly, and at higher temperatures ($\gtrsim 10^7~{\rm K}$) the isochoric modes are stable. We show that even the nonlinear evolution depends on whether the isochoric modes are linearly stable or unstable. For the stable isochoric modes, we observe the growth of small-scale isobaric modes but this is distinct from the nonlinear fragmentation of a dense cooling region. For unstable isochoric perturbations we do not observe large density perturbations at small scales. While very small clouds ($\sim {\rm min}[c_st_{\rm cool}]$) form in the transient state of nonlinear evolution of the stable isochoric thermal instability, most of them merge eventually.
\end{abstract}



\section{Introduction}

The hydrodynamics of optically thin astrophysical plasma under a wide range of physical conditions is  typically modeled using a cooling function ($\Lambda[T,Z]$) that varies with the temperature and the metallicity of the gas (\citealt{Sutherland1993}). 
The temperature dependence of the cooling function determines whether star formation within a galactic halo is efficient or whether star-formation happens via a slow ``cooling flow" (e.g., \citealt{Rees1977}). Similarly, cooling and heating processes have been studied in the context of the structure of the interstellar medium (ISM; \citealt{McKee1977}), 
and the dense solar corona which hosts solar prominences and coronal rain (see \citealt{Antolin2020} for a review).

In this paper, we focus on the multiphase gas in halos of galaxies and clusters, although the physical principles apply broadly. This gas is observed mainly in emission (e.g., \citealt{Fabian2008}) for galaxy clusters  and in absorption in background quasars for galactic halos (e.g., \citealt{Hennawi2015, Lau2016,Muzahid2018}; see also  \citealt{Arrigoni2019}).
Extended, cospatial, warm/ionized, and cold/molecular gas in the cores of galaxy clusters with small volume filling factors (\citealt{Tremblay2018}) implies multiple clumps/filaments dispersed across kpc scales. Broad emission/absorption lines in the context of AGN (commonly explained by a collection of optically thick clouds orbiting around the central engine) have been proposed to be due to "misty" cold gas drifting in the velocity field of a background hot medium (either entrained by fast wind or suspended in the diffuse virialised gas; see section 3 in \citealt{McCourt2018} and references therein). 
But there are also observations of warm gas surrounding  cold gas with no direct indication of mist. Large ${\rm O{VI}}$ columns have been observed in low-redshift Milky Way like galaxies which can be attributed to either large scale cooling flows or spatially extended, low-density photoionized cloud(s) (not necessarily fragmented; see \citealt{McQuinn2018}; see also \citealt{Rudie2019} for high redshift constraints). Thus the observed multiphase gas appears in diverse states, across a wide range of halos and redshifts.

Radiative cooling and heating are thought to play a fundamental role in the multiphase intracluster medium (ICM) of cool-core clusters (e.g., \citealt{Sharma2010,Prasad2015}). Local thermal instability can trigger the formation of the observed dense H$\alpha$ ($\sim 10^4~\rm K$; e.g., \citealt{Fabian2008}) clumps/filaments within the inner hot ICM of dense clusters ({\citealt{Field1965}}, {\citealt{Malagoli1987}},{\citealt{Kim2003}}).
While the temperature dependence of radiative cooling has been studied for local thermal instability in global thermal balance (as inferred in cool cores), the role of metallicity is largely unexplored. This, along with the nonlinear evolution of the multiphase gas, is studied in this paper.

It is well known that the stratification within stars in the mean molecular weight significantly affects the convective stability as expressed in terms of $\partial \ln T/\partial \ln p$ and $\partial \ln \mu/ \partial \ln p$ (\citealt{Ledoux1947,Kato1966}) \footnote{Note that  the buoyancy oscillation frequency is still given by the same expression, $\omega^2 = (k_\perp/k)^2 (g/\gamma) \frac{d}{dr} \ln (p/\rho^\gamma)$, when expressed in terms of the radial entropy gradient.}. Local thermal instability in a gravitationally stratified medium, like cluster cores, appears as linearly overstable buoyancy oscillations (\citealt{Choudhury2016}). Hence, spatial variation of metallicity, mixing and thermal instability mutually affect each other. For example, the variation of helium-to-hydrogen abundance in cluster cores is expected to affect radiative cooling and X-ray properties of cluster cores (\hbox{\citealt{Ettori2006}}, \hbox{\citealt{Peng2009}}, \hbox{\citealt{Berlok2015}}). This also implies that the inclusion of metallicity variation is necessary to understand the nature and evolution of local thermal instability. 


Observations find spatially uniform metallicity profiles in galaxy clusters ($\sim 0.3~Z_{\odot}$, \citealt{Truong2019}, \citealt{Degrandi2001}, \citealt{Molendi2016}) with a small negative radial gradient. 
There is not much evidence for a significant evolution of the global metallicity in clusters for redshifts $\lesssim 1.5$ (\citealt{McDonald2016}). Thus the general picture of metallicity evolution in clusters includes an early enrichment scenario in which most metals are produced by $z\sim 2$. The same enriched gas is ejected, accreted back and thus recycled multiple times. As a result, metallicity remains fairly uniform throughout the medium outside the central core (cool core clusters show a peak in metallicity in the core because of enrichment due to current star formation; \citealt{Leccardi2008,Prasad2018}). Although cluster metallicity is broadly uniform across radius and redshift, stellar winds can locally enrich the surrounding gas
(
e.g., see \citealt{Million2010,Mernier2015}). 
In this work, we find that the metallicity dependence of the cooling function is not as significant for local thermal instability as the temperature dependence. 

The wavelength of isobaric modes is smaller than the scale over which sound waves can propagate in a cooling time ($\sim c_s t_{\rm cool}$). In contrast, the isochoric modes are out of sonic contact over a cooling time. The isobaric thermal instability has been studied in ICM simulations with radiative cooling and various modes of feedback heating (\citealt{McCourt2012}, \citealt{Sharma2012}, \citealt{Li2014}, \citealt{Prasad2015}, \citealt{Choudhury2016}), magnetic fields (\citealt{Ji2018}), rotation (\citealt{Sobacchi2019}), and spatial entropy variation (\citealt{Voit2017}, \citealt{Choudhury2019}). 
Recently \citet{McCourt2018} investigated the fate of an initially isobaric cloud cooling through temperatures $\lesssim 10^6$ K where the cooling time can be shorter than the sound crossing time and cooling is in the isochoric limit. In this case, rather than cooling monolithically, they argue that the cooling cloud should fragment into cloudlets of smaller scale $\sim c_s t_{\rm cool}$ which remain isobaric. The length scale for such an isobaric cloudlet at the peak of the cooling curve is $\lesssim 0.1/n$ pc, much smaller than the global scale of the CGM. The reason for this shattering into tiny cloudlets is a  faster isobaric growth rate compared to an isochoric one. We carry out extensive 1-D nonlinear simulations to test some of these ideas.
\cite{Waters2019} suggest an alternative nonlinear evolution for a growing isochoric cloud that does not shatter. We examine both the isobaric and isochoric regimes in this work and discuss when 
small-scale condensation occurs, for a wide range of background temperatures and metallicities.

The origin of the two end-states (clouds shattering into small cloudlets versus a pulsating monolithic cloud) for isochoric clouds has been discussed in \cite{Gronke2020} using 3D hydrodynamic simulations, in which the initial pressure contrast and overdensity of the cloud govern the  shattering phenomenon. They find that below a final density contrast ($\chi_f = 300$), the 
fragments tend to merge together and the end-state appears to be monolithic. Above the threshold of density contrast, shattering is visually evident in their simulations. Instead of  just classifying the cases of  shattering and no-shattering based on the final state of the cloud, in this paper, we explore the fundamental cause of shattering 
starting with initially linear perturbations ($\delta \rho/\rho < 1$). 
We find that 
the stability (instability) of a linear isochoric mode determines if small scale cold gas will be produced nonlinearly (or not). 

The local slope of the cooling function varies as a function of temperature as different ions dominate radiative losses at various temperatures. The various features in $\partial \ln\Lambda/ \partial \ln T$ can result in ``islands of stability" amid an ``ocean of instability" (\citealt{Pfrommer2013}; see also \citealt{Binney2009}).
We highlight that the "instability/stability islands" differ in the isobaric and isochoric regimes 
and across temperatures. The growth/decay rates
depend not only on the cooling time scale ($t_{\rm cool}$)
but also on the temperature dependence of the cooling function (through $\partial \ln \Lambda/\partial \ln T$). The isochoric modes are thermally stable for $\partial \ln \Lambda/\partial \ln T>0$, and nonlinearly these modes are prone to 
forming clouds separated by small scales, unlike the isochorically unstable modes. 


Our paper is organized as follows. In section \ref{sec:physset}, we describe the physical framework for studying gas evolution with cooling and heating. In section \ref{sec:linstab}, we present the local linear analysis of thermal instability in the isobaric and isochoric regimes. In section \ref{sec:simset}, we describe the simulation setup and in section \ref{sec:res} explain the simulation results. In section \ref{sec:disc}, we discuss our results and conclude.
\section{Physical Setup}
\label{sec:physset}
We start with the simplest model to study the effects of metallicity on the local thermal instability. For the ICM, we take typical density and temperature for the cool core as observed in X-rays. This work does not include gravity but it is well-known that in presence of background gravity, the internal gravity waves generated in the cores will be overstable and will grow due to local thermal instability. Hence, the results obtained in this work can be generalized to an optically-thin, radiatively cooling  gas confined by background gravity. 

The measurement of ICM metallicity and its gradient in observations (\citealt{Degrandi2001},\citealt{Million2010}) motivates us to include a background gradient in our study with a metallicity in the range 0.2$Z_\odot$-0.6$Z_\odot$. 
Mergers, AGN jets (\citealt{Kirkpatrick2015}), stellar winds, ram pressure stripping, etc., can cause the metals to mix beyond the cluster galaxies and even up to the virial radius. 

We evolve a local patch of the intracluster medium in time using the continuity, momentum, and energy equations of hydrodynamics. Additionally, we use tracers to evolve metallicity and H-abundance, and we do not consider any source term for metal or Hydrogen production; pre-existing metals and Hydrogen can be transported by fluid motion. The assumption is that the current rate of metal enrichment is negligible. The hydrodynamics equations that we solve are 
\\
\begin{subequations}
\begin{align}
    \frac{\partial \rho}{\partial t} + \mathbf{\nabla}\cdot\left(\rho\mathbf{v}\right) = 0,  \label{eq:initial1}\\
    \rho \frac{\partial \mathbf{v}}{\partial t} + \rho\left(\mathbf{v}\cdot\mathbf{\nabla}\right)\mathbf{v} = -\mathbf{\nabla}p,  \label{eq:initial2}\\
    \frac{\partial Z}{\partial t} + (\mathbf{v}\cdot\nabla) Z = 0, \label{eq:initial3}\\
        \frac{\partial X}{\partial t} + (\mathbf{v}\cdot\nabla) X = 0, \label{eq:initial3a}\\
     \frac{p}{\gamma-1} \left [ \frac{\partial}{\partial t} + \mathbf{v}\cdot\mathbf{\nabla} \right] \ln\left(\frac{p}{\rho^\gamma}\right) = -q^{-}\left(\rho,T,Z\right) + q^{+},
    \label{eq:initial4}
\end{align}
\end{subequations}\\
where $\rho$ is mass density, $p$ is pressure, $\mathbf{v}$ is velocity, 
$Z$ is metallicity (ratio of metal mass to the total gas mass in a given fluid element), $X$ is the H mass fraction\footnote{The He mass fraction is fixed by $X+Y+Z=1$. The mass fractions $X,~Y,~Z$ satisfy the advection equation, implying that the mass fraction of a Lagrangian fluid element is conserved. Of course, $X$, $Y$, $Z$ can be different for different fluid elements.}, $\gamma=5/3$ is the adiabatic index, $q^{+}$ is the heating rate density and $q^{-}\left(\rho,T,Z\right) = n_i n_e \Lambda\left(T,Z\right)$ ($n_e$ and $n_i$ are electron and ion number densities, respectively) is the cooling rate density. We assume an ideal gas equation of state $p = n k_B T$. In the cooling term, the cooling function $\Lambda(T,Z)$ that we use is given by
\begin{align}
\label{eq:CF}
    &\Lambda\left(T,Z\right) = \Lambda_{\rm H,He}\left(T\right) + \Lambda_{Z,\odot}\left(T\right)\frac{n_e/n_H}{n_{e\odot}/n_{H\odot}}\frac{Z}{Z_\odot}, 
\end{align}
where 
$\Lambda_{\rm H,He}$ and $\Lambda_{Z,\odot}n_e n_{H\odot}/\left(n_{e\odot}n_H\right)$ for different values of temperature between $100~\rm K$ and $9.6 \times 10^8~\rm K$ are used from tables of \citet{Wiersma2009}. 

Since cool cores do not undergo catastrophic cooling, we include a simple heating function that imposes global thermal balance, crudely mimicking AGN feedback. This heating scheme is not dependent on any state-variable (namely density, temperature) and is well motivated for the ICM where the dissipation of mechanical energy is the dominant heating source. 
The role of this heating function in our model is to simply maintain 
the observed rough global thermal balance. Therefore, the time-independent heating rate density ($q^+[{\bf r}]$) is
simply equal to the cooling rate density in the unperturbed background with a given density, temperature and metallicity profile. Note that this heating scheme often causes overheating in diffuse gas at late times. Such unphysically hot/dilute gas will be absent with stratification, where hot gas can expand and cool adibatically.

\section{Linear Stability Analysis}
\label{sec:linstab}
We perform a linear  WKB analysis for the local model of a static intracluster medium, with fluctuations in all variables including the metallicity and H-abundance. For simplicity, we assume the background density and pressure to be uniform, and the background velocity to be zero. However, we allow a gradient
in the background metallicity and H-abundance. We consider plane wave perturbations varying as $e^{i(kx-\omega t)}$ (i.e., $x-$ axis is chosen along the wavenumber direction).

Linearizing the hydrodynamical equations (excluding the energy equation), we get
\begin{subequations}
\begin{align}
	- \omega \delta \rho + k\rho_0 \delta v_x  = 0, \label{eq:continuity}\\ 
	\rho_0 \omega \delta v_x - k \delta p = 0, \label{eq:momentum}\\ 
	i\omega \delta Z - \delta v_x \frac{\partial Z_0}{\partial x} = 0, \label{eq:dye}\\
	i\omega \delta X - \delta v_x \frac{\partial X_0}{\partial x} = 0, \label{eq:dye_a}
\end{align}
\end{subequations}
and velocity fluctuations perpendicular to ${\bf k}$ vanish.

\noindent Let $X$, $Y$ and $Z$ be the mass fractions of H, He and metals (mostly O). Assuming a fully ionized plasma, we can relate the electron and ion number densities to these as follows,
\begin{subequations}
\begin{align}
\label{eq:ni}
	n_i = \frac{\rho}{\mu_i m_p} = \frac{\rho}{m_p}\left(X + \frac{Y}{4} + \frac{Z}{16}\right), \\
\label{eq:ne}
	n_e = \frac{\rho}{\mu_e m_p} = \frac{\rho}{m_p}\left(X + \frac{Y}{2} + \frac{Z}{2}\right).
\end{align}
\end{subequations}
From Eqs. \ref{eq:ni}, \ref{eq:ne}, we get for the perturbed number densities
\begin{subequations}
\begin{align}
\label{eq:dni}
\delta n_i &= \frac{\delta\rho}{\rho_0}n_{i0} + \frac{\rho_0}{m_p}\left(\frac{3}{4}\delta X - \frac{3}{16} \delta Z \right),~{\rm and} \\
\label{eq:dne}
    \delta n_e &= \frac{\delta\rho}{\rho_0}n_{e0} +
    \frac{\rho_0}{m_p}\left(\frac{1}{2}\delta X \right),
    \end{align}
\end{subequations}
where we have used $\delta X + \delta Y + \delta Z=0$ to eliminate $\delta Y$.

\noindent Linearizing $q^-$ on the right hand side of Eq.~\ref{eq:initial4}, we get
$$
	\frac{\delta q^{-}}{q_0^-} =  \Lambda_T\frac{\delta T}{T_0} + \Lambda_Z\frac{\delta Z}{Z_0} + \frac{\delta n_i}{n_{i0}} + \frac{\delta n_e}{n_{e0}},
$$	
where 
$\Lambda_T \equiv (\partial \ln \Lambda/\partial  \ln T)_0$,
$\Lambda_Z \equiv (\partial \ln \Lambda/\partial \ln Z)_0$, and  $q_0^{-} = n_{i0}n_{e0} \Lambda_0$.

\noindent Therefore, the linearized right hand side of the energy equation (Eq. \ref{eq:initial4}) becomes
$$
 -\delta q^- + \delta q^+ = - q_0^{-}\left( \Lambda_T\frac{\delta T}{T_0} + \Lambda_Z\frac{\delta Z}{Z_0} + \frac{\delta n_i}{n_{i0}} + \frac{\delta n_e}{n_{e0}} \right) + \delta q^{+}.
$$
In the background equilibrium state, $q^+_0 = q^-_0$.
And as mentioned earlier, for simplicity, 
we assume that $q^+$ depends only on position, 
so its Eulerian perturbation vanishes; i.e., $\delta q^+=0$.
 Note that as long as there is global thermal balance, we expect a similar evolution and saturation of local thermal instability even if there are fluctuations in the heating rate (see section 5.4 in \citealt{McCourt2012}). Of course, the thermal instability growth rate, in general, depends on the assumed dependence of the heating rate density ($q^+$) on density, temperature and metallicity (here we assume $\partial q^+/\partial T = \partial q^+/\partial \rho = \partial q^+/\partial Z=0$; \citealt{McCourt2012}).

Using the expressions for $\delta n_e$ and $\delta n_i$ (Eqs. \ref{eq:dne}, \ref{eq:dni}), the right hand side of the energy equation (Eq. \ref{eq:initial4}) becomes
$$
	 -q_0^{-}\left[ \Lambda_T\frac{\delta T}{T_0} + \Lambda_Z\frac{\delta Z}{Z_0} + \frac{2\delta\rho}{\rho_0} + \frac{\rho_0}{m_p n_{i0}} \left( \left\{\frac{3}{4} + \frac{n_{i0}}{2n_{e0}}\right\} \delta X - \frac{3}{16} \delta Z \right) \right].
$$

After equating the left and right hand sides of the linearized energy equation (Eq. \ref{eq:initial4}), we have
\begin{align}
    -i\omega\left[ \frac{\delta p}{p_0} - \frac{\gamma \delta \rho}{\rho_0}  \right] = -\frac{\left(\gamma -1\right)q_0^{-}}{p_0}\left[ \Lambda_T\frac{\delta T}{T_0}  + \frac{2\delta\rho}{\rho_0} + \alpha\delta Z + \beta \delta X \right], \nonumber
\end{align}
where $\alpha =  \Lambda_Z/Z_0 -3 \rho_0/16 m_p n_{i0}$ and $\beta = (3/4 + n_{i0}/2n_{e0})(\rho_0/m_pn_{i0})$.

After introducing an equilibrium cooling time $t_{\rm cool,0} \equiv p_0/[\left(\gamma -1\right)q_0^{-}]$, and expressing temperature fluctuations in terms of pressure, density and metallicity fluctuations, the linearized energy equation becomes
\begin{align}
\label{eq:delp}
    &-i\omega\left[ \frac{\delta p}{p_0} - \frac{\gamma \delta \rho}{\rho_0}  \right] = -\frac{1}{t_{\rm cool, 0}}\left[ \Lambda_T\frac{\delta p}{p_0}  + \frac{\left(2 - \Lambda_T\right) \delta\rho}{\rho_0} + \epsilon \delta Z + \eta \delta X \right],
\end{align}
where $\epsilon = \alpha + 3 \rho_0\Lambda_T/(16 m_p n_{0}) $, $\eta  = \beta + 5 \rho_0\Lambda_T/(4 m_p n_{0})$ and $n_0 = n_{i0} + n_{e0}$.

From Eqs. \ref{eq:continuity}-\ref{eq:dye_a} we get $k^2 \delta p = \omega^2 \delta \rho$, $\delta Z = -i \delta \rho (\partial Z_0/\partial x)/k \rho_0$, and $\delta X = -i \delta \rho (\partial X_0/\partial x)/k \rho_0$.
Using expression for $\delta Z$, rearranging the terms, and defining 
\begin{equation}
    t_{\rm TI} =  \frac{\gamma t_{\rm cool,0}}{\left(2-\Lambda_T\right)},
\label{eq:tTI}
\end{equation}
we have

\begin{align}
    \left(i\omega +  \frac{1}{t_{\rm TI}} - \frac{i}{\gamma k t_{\rm cool,0}} \left[ \epsilon \frac{\partial Z_0}{\partial x} +  \eta \frac{\partial X_0}{\partial x} \right] \right) & \frac{\delta \rho}{\rho_0} \nonumber \\
    = & \left( \frac{i\omega}{\gamma} - \frac{\Lambda_T}{\gamma t_{\rm cool,0}} \right) \frac{\delta p}{p_0}. \label{delpdelrho}
\end{align}

Eqs. \ref{eq:continuity}-\ref{eq:dye_a}, \ref{eq:delp} are five linear equations and have five roots in general, two corresponding to the oppositely-traveling sound waves and three
entropy modes. Of course, all these modes are modified by cooling and heating. Using the relation between $\delta\rho$ and $\delta p$ in Eq. \ref{delpdelrho} and $\omega^2\delta \rho = k^2 \delta p$, we get the dispersion relation
\begin{eqnarray}
    \omega^3 + \frac{i\Lambda_T}{t_{\rm cool,0}} \omega^2 + c_{\rm s}^2 k^2 \omega + \frac{c_{\rm s}^2 k}{\gamma k t_{\rm cool,0}} \left[ H^{-1} + ik(2-\Lambda_T) \right] = 0. \label{eq:cubic}
\end{eqnarray}
where $H^{-1}$ = $[\epsilon (\partial Z_0/\partial x) + \eta (\partial X_0/\partial x) ]$. Eq. \ref{eq:cubic} is a cubic equation, which has three roots corresponding to two sound waves and one entropy mode. These three roots are three out of the five modes expected from Eqs. \ref{eq:continuity}-\ref{eq:dye_a}. The remaining two modes are trivial ($\omega =0$ with $\delta X,~\delta Y \neq 0$) modes. We can simplify the linear analysis for the condensation thermal instability modes in the limit of fast and slow sound-crossing times. The sound waves are generally damped (unlike the entropy mode, which grows) for typical cooling curves (\citealt{Field1965}), so we focus on the condensation modes.

Note that
$$
    \frac{ \delta p/p_0}{\delta \rho/\rho_0} = \frac{\omega^2}{k^2} \frac{\gamma}{c_s^2} \sim \frac{\gamma t_{\rm snd}^2}{t_{\rm cool}^2},
$$
where $t_{\rm snd}^2 = 1/(k^2 c_s^2)$ ($t_{\rm snd}$ is the sound crossing time across the mode), $c_s^2 = \gamma p_0/\rho_0$, and we assume the modes to evolve on the cooling time ($\omega \sim t_{\rm cool}^{-1}$).

In the isobaric limit, $t_{\rm snd} \ll t_{\rm cool}$ and $\delta p/p_0 \ll \delta \rho/\rho_0$, the dispersion relation becomes
\begin{equation}
\label{eq:isobaric}
    \omega =  \frac{i}{t_{\rm TI}} + \frac{1}{\gamma k t_{\rm cool,0}} \left( \epsilon \frac{\partial Z_0}{\partial x} + \eta \frac{\partial X_0}{\partial x} \right),
\end{equation}
where $t_{\rm TI}$ is given by Eq. \ref{eq:tTI}.
While in the isochoric limit, $t_{\rm snd} \gg t_{\rm cool}$ and $\delta p/p_0 \gg \delta \rho/\rho_0$, the dispersion relation is
\begin{equation}
\label{eq:isochoric}
\omega =  \frac{-i\Lambda_T}{t_{\rm cool, 0}}.
\end{equation}

In the isobaric limit, valid for small wavelength modes, we get an oscillatory response in the presence of metallicity gradient. We note that the growth rate in the isobaric and isochoric (valid for large scale modes) limits are identical to the expressions in absence of metallicity gradients. Overall, the impact of metallicity gradient is expected to be modest. This result is anticipated physically because the metallicity gradient is rather small and the Lagrangian derivative of metallicity vanishes (Eq. \ref{eq:initial3}), implying that the metallicity (unlike temperature/density) is comparable in over- and under-dense regions.
\begin{figure}
    \includegraphics[width=\figsize\textwidth]{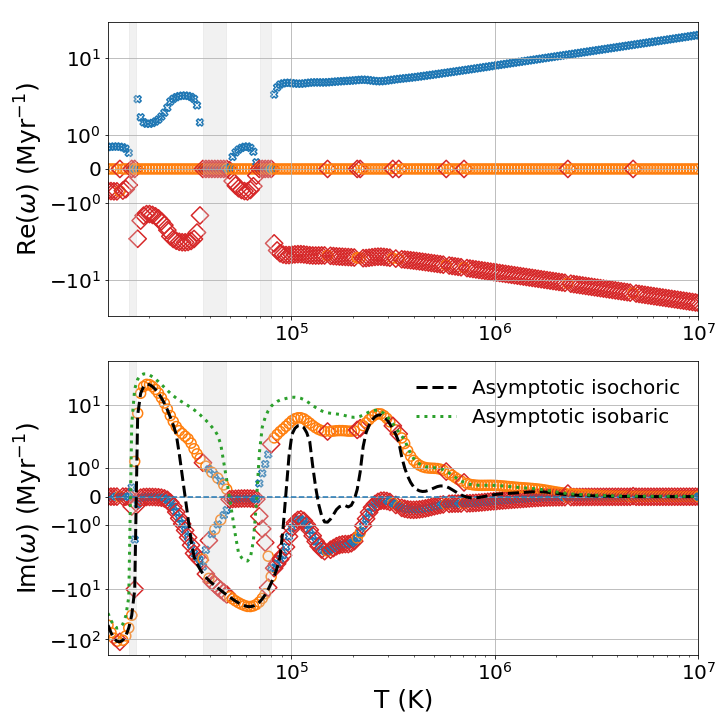}
    \centering
    \caption{Real and imaginary parts of solutions of Eq. \ref{eq:cubic} at different temperatures with $H^{-1}=0$, $Z_0=0.6 Z_\odot$, $X_0=1.01X_\odot$ and $k=2\pi/(0.1 \rm kpc)$. The different markers and their colors are consistent among the two plots and can be used to distinguish the modes (blue crosses and red diamonds usually correspond to the stable sound waves and orange circles to the unstable entropy mode). The shaded regions are temperatures where the distinction between the sound and entropy modes is not clear. Green dotted and black dashed lines are the asymptotic isobaric and isochoric growth rates from Eqs. \ref{eq:isobaric} \& \ref{eq:isochoric}.}
    \label{fig:cubic_intermediate}
\end{figure}

Fig. \ref{fig:cubic_intermediate} shows the real and imaginary parts of the three non-trivial roots of Eq. \ref{eq:cubic} for an intermediate (neither isobaric nor isochoric) value of $k$. We use $H^{-1}=0$, $Z_0=0.6 Z_\odot$, $X_0=1.01X_\odot$ and $k=2\pi/(0.1 \rm kpc)$ to calculate these roots. There are some regions in temperature (shown as shaded regions in Fig. \ref{fig:cubic_intermediate}) where the three modes couple. Outside these regions, the root with zero real part corresponds to the entropy mode and the other two roots correspond to sound waves. Also, the growth rate of the entropy mode for an intermediate $k$ lies between the asymptotic isochoric and the asymptotic isobaric growth rates.

From Eq. \ref{eq:isobaric} and \ref{eq:isochoric}, we see that there are three possible regimes depending on the value of $\Lambda_T$. The regime $\Lambda_T < 0$ corresponds to unstable isobaric and isochoric modes, $0<\Lambda_T<2$ corresponds to unstable isobaric but stable isochoric modes, and $\Lambda_T>2$ corresponds to stable isobaric and isochoric modes. In section \ref{sec:nonlinear}, we highlight the differences in the nonlinear evolution in the first two of these regimes relevant for the CGM and cool core clusters, respectively. We emphasize that there is no explicit dependence of the thermal instability growth rate on the metallicity via $\Lambda_Z$ (unlike $\Lambda_T$) in both isobaric and isochoric regimes.
\section{Simulation setup}
\label{sec:simset}

\begin{figure}
    \includegraphics[width=\figsize\textwidth]{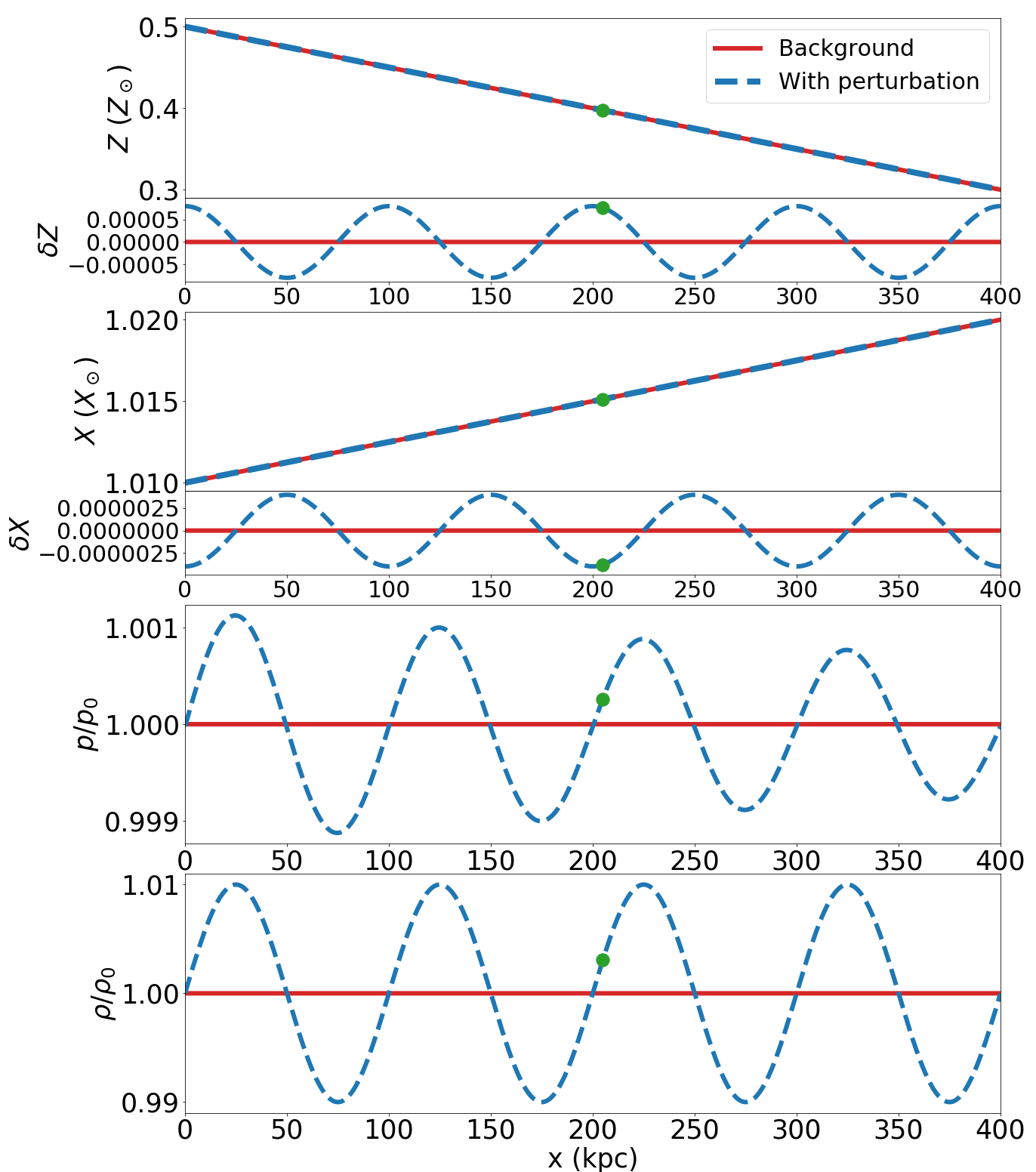}
    \centering
    \caption{Isobaric initial condition with $k=2\pi/(100~{\rm kpc})$. The top panel shows the metallicity, the second panel H-abundance, the third panel pressure, and the lowest panel density as a function of radius. Both the background without and with perturbations are shown. The perturbations are shown separately for the metallicity and H-abundance which have a background linear gradient. The green dot indicates the location used for measuring the growth and oscillation rates in section \ref{sec:compwsim}.}
    \label{fig:initial}
\end{figure}

We use the {\tt PLUTO} code (\citealt{Mignone2007}) to solve the hydrodynamical equations (Eq. \ref{eq:initial1}-\ref{eq:initial4}) on a one-dimensional (1-D) grid, starting from a given initial condition at $t=0$. We carry out two broad classes of simulations: ones to investigate linear evolution (see Table \ref{tab:linear_sim} \& section \ref{sec:compwsim}), and others to study nonlinear evolution (see Table \ref{tab:nonlinear_sim} \& section \ref{sec:nonlinear}).

\subsection{Initial Conditions}
For both the linear and nonlinear simulations, the unperturbed background density ($\rho_0 \approx 0.062 m_p $ g cm$^{-3}$; $m_p$ is proton mass) and pressure (varies for different runs) are uniform across the entire grid. In some of our simulations, we allow the metallicity to have a mild gradient. 

We seed fluctuations in all dependent variables, using eigenmodes from our linear analysis in section \ref{sec:linstab}. For a given wavenumber $k$, the frequency is obtained from the asymptotic dispersion relations in Eq. \ref{eq:isobaric} or Eq. \ref{eq:isochoric}, depending on whether $t_{\rm cool} >$ or $< 1/(kc_s)$, the isobaric and isochoric limits, respectively. We use the frequency obtained in Eqs. \ref{eq:continuity}-\ref{eq:dye_a} 
to calculate the perturbations in terms of the amplitude $\delta\rho$ and wavenumber $k$. Fig.~\ref{fig:initial} shows an example of an isobaric initial condition used in our simulations.

\subsection{Grid and boundary conditions}
To study the linear evolution (see section \ref{sec:compwsim}), we run the simulations with 4096 grid points with different box sizes, depending on the value of the wavenumber ($k$). We run our simulations in the isobaric and isochoric regimes, with a corresponding value of $k$. 
All simulations used to compare with the linear theory are listed in Table \ref{tab:linear_sim}. 

\begin{table*}
    \caption{Simulations to compare with linear theory}
	\label{tab:linear_sim}
    \begin{tabular}{cccccccccccc}
    \hline
    \multicolumn{1}{|p{1cm}|}{\centering Simulation\\ID$^\dag$} & 
    \multicolumn{1}{|p{1.5cm}|}{\centering $k$\\($\rm kpc^{-1}$)} & 
    \multicolumn{1}{|p{1cm}|}{\centering Limit} & 
    \multicolumn{1}{|p{1cm}|}{\centering $\dfrac{dZ_0}{dx}$\\$(Z_\odot/\rm kpc$)} & 
    \multicolumn{1}{|p{1cm}|}{\centering $Z_{0~(\rm max)}$\\$(Z_\odot)$} &
    \multicolumn{1}{|p{1cm}|}{\centering $\dfrac{dX_0}{dx}$\\$(X_\odot/\rm kpc$)$^\ddag$} & 
    \multicolumn{1}{|p{1cm}|}{\centering Box size\\(kpc)} &
    \multicolumn{1}{|p{1.5cm}|}{\centering Initial \\$T_{\rm avg}$ ($10^6~K$)} & 
    \multicolumn{1}{|p{1.5cm}|}{\centering ${\rm Im}\left(\omega\right)$\\(Theory)\\(Myr$^{-1}$)} & 
    \multicolumn{1}{|p{1.5cm}|}{\centering ${\rm Im}\left(\omega\right)$\\(Simulations)\\(Myr$^{-1}$)}\\
    \hline
    IB\_L2kpc & $\pi$ & Isobaric& $-2.5\times10^{-2}$ & 0.5 & $2.5\times10^{-5}$ & 8 & 8.89 & 0.00677 & 0.00677  \\
    IB\_L4kpc & $0.5\pi$ & Isobaric & $-5\times10^{-3}$ & 0.5 & $2.5\times10^{-4}$ & 40 & 8.85 & 0.00683 & 0.00666 \\
    IB\_L20kpc & $0.1\pi$ & Isobaric & $-5\times10^{-4}$ & 0.5 & $2.5\times10^{-5}$ & 400 & 8.85 & 0.00683 & 0.00708 \\
    \hline
    IC\_L2Mpc & $0.001\pi$ & Isochoric & $-2.5\times10^{-5}$ & 0.5 & $1.25\times10^{-6}$ & 8000 & 8.85 & -0.00075 & -0.00073 \\
    IC\_L4Mpc & $0.0005\pi$ & Isochoric & $-5\times10^{-6}$ & 0.5 & $2.5\times10^{-7}$ & 40000 & 8.85 &  -0.00075 & -0.00079 \\
    IC\_L20Mpc & $0.0001\pi$ & Isochoric & $-2.5\times10^{-6}$ & 0.5 & $1.25\times10^{-7}$ & 60000 & 8.74 & -0.00072 & -0.00079\\
    \hline
    IC\_L200kpc & $0.01\pi$ & Isochoric & $-5\times10^{-4}$ & 0.5 & $2.5\times10^{-5}$ & 400 & 8.85 &  -0.00075 & 0.00157 \\
    IC\_L100kpc & $0.02\pi$ & Isochoric & $-5\times10^{-4}$ & 0.5 & $2.5\times10^{-5}$ & 400 & 8.85 & 0.00075 & 0.00469  \\
    IB\_L100kpc & $0.02\pi$ & Isobaric & $-5\times10^{-4}$ & 0.5 & $2.5\times10^{-5}$ & 400 & 8.85 &  0.00683 & 0.00278 \\
    \hline
    F\_L20kpc & $0.1\pi$ & Isobaric & $-5\times10^{-5}$ & 0.59 & $2.5\times10^{-6}$ & 400 & 31.08 &  0.00191 & 0.00198 \\
    LT\_L20kpc & $0.1\pi$ & Isobaric & $-5\times10^{-5}$ & 0.3 & $2.5\times10^{-6}$ & 400 & 27.55 &  0.00192 & 0.00202 \\
    LZ\_L20kpc & $0.1\pi$ & Isobaric & $-5\times10^{-5}$ & 0.35 & $2.5\times10^{-6}$ & 400 & 22.21  & 0.00246 & 0.00256 \\
    \hline
    F\_L2Mpc & $0.001\pi$ & Isochoric & $-2.5\times10^{-6}$ & 0.59 & $2.5\times10^{-7}$ & 6000 & 31.07 &  -0.00056 & -0.00055 \\
    LT\_L2Mpc & $0.001\pi$ & Isochoric & $-2.5\times10^{-6}$  & 0.35 & $2.5\times10^{-7}$ & 6000 & 27.54 &  -0.00055 & -0.00053 \\
    LZ\_L2Mpc & $0.001\pi$ & Isochoric & $-2.5\times10^{-6}$  & 0.3 & $2.5\times10^{-7}$ & 6000 & 22.21 & -0.00041 & -0.00039 \\
    \hline
    \end{tabular}
    \\
    $^\dag$ IB : Isobaric initial conditions, IC : Isochoric initial conditions, F : Fiducial case, LT : Same $\Lambda_T$ as fiducial case, LZ : Same $\Lambda_Z$ as fiducial case,
    \\ L100kpc : Denotes the wavelength of the initial perturbations, in this case 100 kpc. $^\ddag$ $X_{0~(\rm min)}=1.01X_\odot$ for all simulations in this table.\\
    \textbf{Notes:} These simulations have a background density of 0.062$m_p$ g cm$^{-3}$, initial $\delta \rho/\rho = 0.01$, and 4096 grid cells in the simulation box. The growth rate measured in simulations for intermediate $k$s does not match the asymptotic theory (see Fig. \ref{fig:growth}).
\end{table*}

To study nonlinear evolution (see section \ref{sec:nonlinear}), we run simulations with isochoric initial conditions for $t>20t_{\rm cool}$ and those with isobaric initial conditions for $t\approx5t_{\rm cool}$. We run the isobaric simulations for a shorter duration as the isobaric instability grows faster and leads to very high temperatures in the diffuse regions (because of the imposed global thermal balance), which make the code timestep very small. However, even our isobaric runs are deep into the nonlinear regime. 

For the nonlinear runs, we use simulation boxes of different sizes for isobaric and isochoric simulations. All nonlinear simulations (listed in Table \ref{tab:nonlinear_sim}) use 8192 grid points (except the high resolution [HR] simulations, that use 20000 grid points) . The size of the simulation box for the isochoric simulations is 6 Mpc, which is equal to 3 wavelengths of the mode with wavenumber $k=\pi/(10^{3}~\rm kpc)$; these wavelengths are unrealistically large but are carried out to understand the evolution of isochoric modes. In reality, the isochoric evolution occurs in the transient nonlinear phase when temperature is close to the peak of the cooling curve ($\sim 10^5$ K), but even the nonlinear evolution seems to be primarily governed by linear physics (c. f. section \ref{sec:frag_isochoric}). On the other hand, the simulation box for isobaric simulations extends 400 kpc, which can fit 20 wavelengths of the mode with $k=\pi/(10~\rm kpc)$. 

\begin{figure}
    \includegraphics[width=\figsize\textwidth]{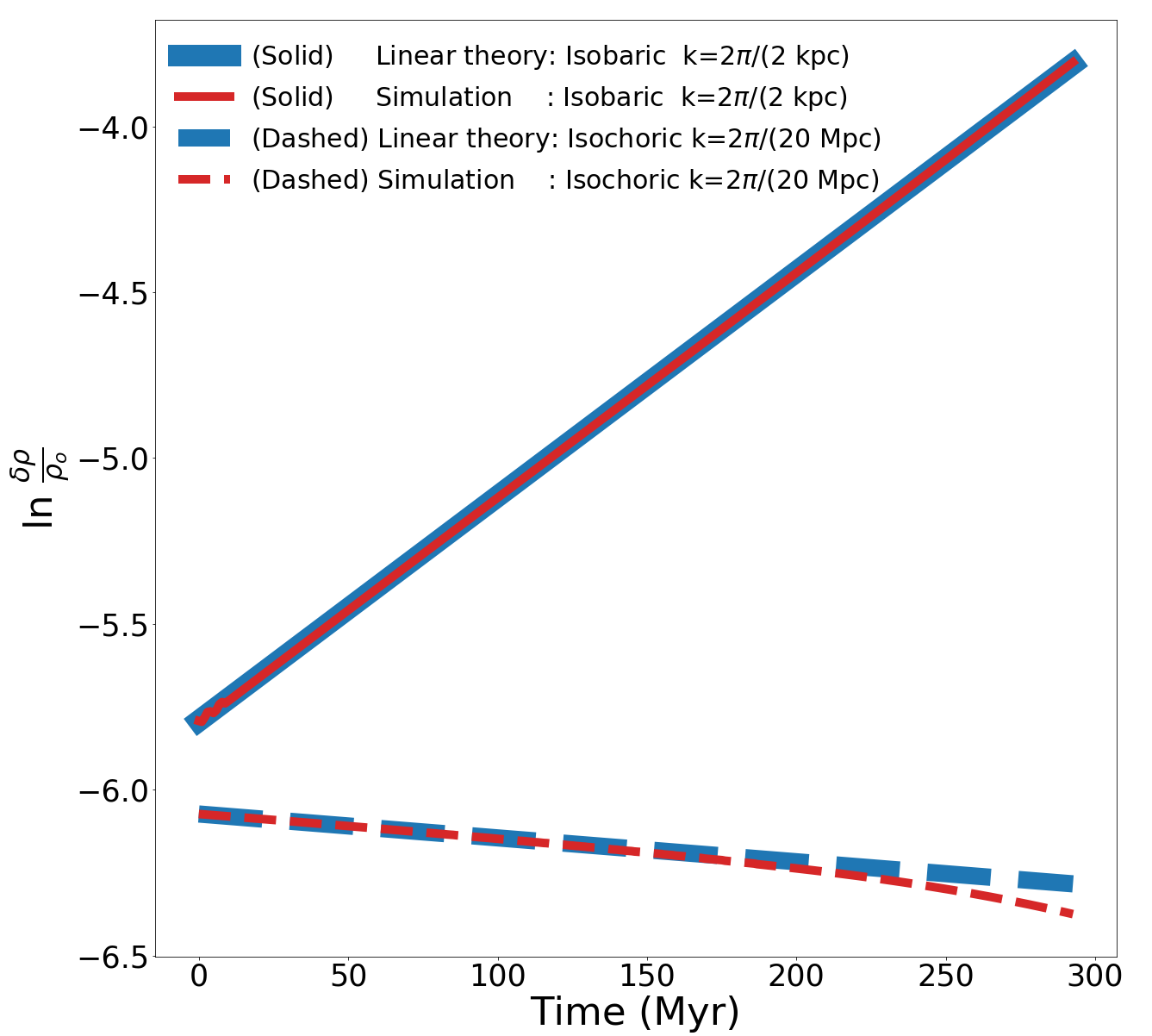}
    \centering
    \caption{Comparison of the growth rate in simulations and linear theory for $k=2\pi/(2~{\rm kpc})$ (isobaric, solid lines) and $2 \pi/(20~{\rm Mpc})$ (isochoric, dashed lines). For exact comparison of the slope, we choose the same initial amplitude for the simulation and the analytic eigenmode. }
    \label{fig:rhoisobaric}
\end{figure}
For simulations with no metallicity gradient, we use periodic boundary conditions. With metallicity gradient, we use outflow boundary conditions at both boundaries and have a buffer region near boundaries where we turn off heating and cooling. We put these buffer regions to avoid condensation and heating near boundaries. 
Buffer zones are not needed with periodic boundaries as the background is translationally invariant.
We set the cooling function ($\Lambda[T,Z]$) to zero below $10^4~\rm K$ and above $9\times 10^8~\rm K$ to prevent the formation of extremely cold and hot regions. We also impose a temperature floor at $10^4~\rm K$.

\begin{table*}
    \caption{Nonlinear simulations}
	\label{tab:nonlinear_sim}
    \begin{tabular}{cccccccccc}
    \hline
    \multicolumn{1}{|p{1.5cm}|}{\centering Simulation\\ID$^\dag$} &
    \multicolumn{1}{|p{1cm}|}{\centering $k$\\($\rm kpc^{-1}$)} & 
    \multicolumn{1}{|p{1.5cm}|}{\centering Limit\\} &
    \multicolumn{1}{|p{1.5cm}|}{\centering Initial\\conditions} &
    \multicolumn{1}{|p{1cm}|}{\centering Boundary\\condition} & 
    \multicolumn{1}{|p{1cm}|}{\centering $\dfrac{dZ_0}{dx}$\\$(Z_\odot/\rm kpc$)$^\ddag$}& 
    \multicolumn{1}{|p{1cm}|}{\centering $\dfrac{dX_0}{dx}$\\$(X_\odot/\rm kpc$)$^\ddag$}&
    \multicolumn{1}{|p{1cm}|}{\centering Box size\\(kpc) } &
    \multicolumn{1}{|p{2cm}|}{\centering Initial\\$T_{\rm avg}$ ($10^6$ K)} & \multicolumn{1}{|p{1cm}|}{\centering $t_{\rm cool}$\\(Myr) } \\ 
    
    \hline
    NIC\_ST\_P & $0.001\pi$ & Isochoric & \multicolumn{1}{|p{1.5cm}|}{\centering Single\\eigenmode} & Periodic & $0$ & $0$ & 6000 & 6.84 & 109.71   \\
    NIC\_UST\_P & $0.001\pi$ & Isochoric & \multicolumn{1}{|p{1.5cm}|}{\centering Single\\eigenmode} & Periodic & $0$ & $0$ & 6000 & 1.95 & 11.79  \\
    NIC\_P\_LOWT & $0.001\pi$ & Isochoric & \multicolumn{1}{|p{1.5cm}|}{\centering Single\\eigenmode} & Periodic & $0$ & $0$ & 6000 & 0.18 & 0.19   \\
    \hline
    NIB\_OF & $0.1\pi$ & Isobaric &\multicolumn{1}{|p{1.5cm}|}{\centering Single\\eigenmode} & Outflow & $-5\times10^{-4}$ & $2.5\times10^{-5}$ & 400 & 8.85 & 134.26  \\
    NIC\_ST\_OF & $0.001\pi$ & Isochoric & \multicolumn{1}{|p{1.5cm}|}{\centering Single\\eigenmode} & Outflow & $-5\times10^{-5}$ & $2.5\times10^{-6}$ & 6000 & 6.80 & 109.22   \\
    NIC\_UST\_OF & $0.001\pi$ & Isochoric & \multicolumn{1}{|p{1.5cm}|}{\centering Single\\eigenmode} & Outflow & $-5\times10^{-5}$ & $2.5\times10^{-6}$ & 6000 & 1.98 & 12.32  \\
    \hline
    NIC\_ST\_ME & $\pi$,~$0.001\pi$ & \multicolumn{1}{|p{1.5cm}|}{\centering Isobaric\\+ Isochoric} & \multicolumn{1}{|p{1.5cm}|}{\centering Mixed\\eigenmode} & Outflow & $-5\times10^{-5}$ & $2.5\times10^{-6}$ & 6000 & 6.80 & 109.22 \\
    NIC\_UST\_ME & $\pi$,~$0.001\pi$ & \multicolumn{1}{|p{1.5cm}|}{\centering Isobaric\\+ Isochoric} & \multicolumn{1}{|p{1.5cm}|}{\centering Mixed\\eigenmode} & Outflow & $-5\times10^{-5}$ & $2.5\times10^{-6}$ & 6000 & 1.98 & 12.32  \\
    \hline
    NIC\_ST\_RN & $0.001\pi$ & Isochoric & \multicolumn{1}{|p{1.5cm}|}{\centering Random\\noise} & Outflow & $-5\times10^{-5}$ & $2.5\times10^{-6}$ & 6000 & 6.80 & 109.22   \\
    NIC\_UST\_RN & $0.001\pi$ & Isochoric & \multicolumn{1}{|p{1.5cm}|}{\centering Random\\noise} & Outflow & $-5\times10^{-5}$ & $2.5\times10^{-6}$ & 6000 & 2.27 & 18.02  \\
    \hline
    NIC\_ST\_HR & $2000\pi$,~$20\pi$ & \multicolumn{1}{|p{1.5cm}|}{\centering Isobaric\\+ Isochoric} & \multicolumn{1}{|p{1.5cm}|}{\centering Mixed\\eigenmode} & Periodic & $0$ & $0$ & 0.1 & 0.04 & 0.14   \\
    NIC\_UST\_HR & $2000\pi$,~$20\pi$ & \multicolumn{1}{|p{1.5cm}|}{\centering Isobaric\\+ Isochoric} & \multicolumn{1}{|p{1.5cm}|}{\centering Mixed\\eigenmode} & Periodic & $0$ & $0$ & 0.1 & 0.30 & 0.45  \\
    \hline
    NIC\_UST\_SQ & NA & Isochoric & \multicolumn{1}{|p{1.5cm}|}{\centering Nonlinear\\ square wave} & Periodic & $0$ & $0$ & 400 & \multicolumn{1}{|p{2cm}|}{\centering 10 (Overdensity)\\20 (Ambient)} & 75.01  \\
    NIC\_ST\_SQ & NA & \multicolumn{1}{|p{1.5cm}|}{\centering Isochoric} & \multicolumn{1}{|p{1.5cm}|}{\centering Nonlinear \\ square wave} & Periodic & $0$ & $0$ & 2000 & \multicolumn{1}{|p{2cm}|}{\centering 30 (Overdensity)\\60 (Ambient)} & 259.39  \\
    \hline
    \end{tabular}
    $^\dag$ NIB : Nonlinear isobaric simulation, NIC : Nonlinear isochoric simulation, ST : Stable isochoric mode, UST : Unstable isochoric mode,\\
    P : Periodic boundary condition, OF : Outflow boundary condition, ME : Mixed eigenmode initial conditions, RN : Random noise initial conditions,\\ 
    LOWT : Low temperature simulation ($T\sim 10^5 \rm K$), HR : High resolution simulations.
    $^\ddag$ $Z_{0~(\rm max)}=0.6Z_\odot$, $X_{0~(\rm min)}=1.01X_\odot$ for all simulations in this table. \\ \textbf{Note:} All simulations have $\rho_0 = 0.062 m_p$ g cm$^{-3}$, $\delta \rho/\rho_0=10^{-5}$, and 8192 grid points (except HR which has $\delta \rho/\rho_0=10^{-4}$ and 20000 grid points; also, SQ runs use a high amplitude square wave density perturbation).
\end{table*}

\section{Simulation results}
\label{sec:res}
\subsection{Comparison of linear theory and simulations}
\label{sec:compwsim}

We compare the linear growth/decay rates of the modes in the isobaric/isochoric regimes ($k>2\pi/[100~{\rm kpc}],~k<2 \pi/[200~\rm kpc]$ respectively, at T$\approx9\times10^6$ K) with the initial evolution in our simulations. 
From simulations, we calculate $\delta\rho/\rho_0$ at a fixed point in space (indicated in Fig.~\ref{fig:initial}) as $\left(\rho-\rho_0\right)/\rho_0$, where $\rho_0$ is the background density. For theoretical prediction, we calculate the corresponding growth rate from linear theory (Eqs. \ref{eq:isobaric}, \ref{eq:isochoric}) at the same point using the background density, temperature and metallicity. Fig.~\ref{fig:rhoisobaric} shows the normalized density perturbation amplitude as a function of time. In the initial stage, the growth and decay rates of the isobaric and isochoric modes match well with linear theory.
\begin{figure}
    \includegraphics[width=\figsize\textwidth]{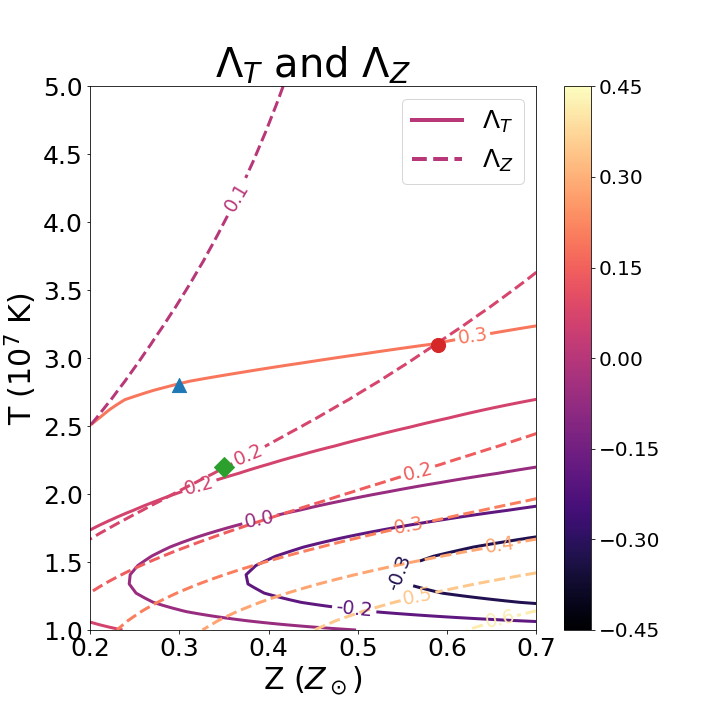}
    \centering
    \caption{Contours of constant $\Lambda_T$ (solid lines) and $\Lambda_Z$ (dashed lines) shown for different temperatures and metallicities with the values determined by the colormap. The red circle (the fiducial case), blue triangle (with the same $\Lambda_T$), and green diamond (with the same $\Lambda_Z$) markers represent the three simulations mentioned in section \ref{sec:compwsim} used to demonstrate the
    dependence (independence) of the linear growth rate on $\Lambda_T$ ($\Lambda_Z$). }
    \label{fig:lamcont}
\end{figure}

For further validation of linear theory, we check the dependence of the linear growth rate on important parameters $\Lambda_T \equiv \partial \ln \Lambda/\partial \ln T$ and $\Lambda_Z \equiv \partial \ln \Lambda/\partial \ln Z$. According to section~\ref{sec:linstab}, the linear growth rate depends explicitly on $\Lambda_T$ but not on  $\Lambda_Z$. Fig.~\ref{fig:lamcont} shows the contours of $\Lambda_T$ and $\Lambda_Z$ in the $T-Z$ plane. We choose three combinations of temperature and metallicity (represented by red circle, blue triangle and green diamond markers) for our runs to check the dependence of the linear growth rate on $\Lambda_T$ and $\Lambda_Z$. The red circle lies on the intersection of contours of $\Lambda_T=0.3$ and $\Lambda_Z=0.2$, while the other two have a different $\Lambda_Z$ (blue triangle) and $\Lambda_T$ (green diamond) compared to the intersection point. These combinations of $(T,Z)$  result in two pairs of systems. One pair has the same $\Lambda_T$ but different $\Lambda_Z$ and the other pair has the same $\Lambda_Z$ but different $\Lambda_T$. 

\begin{figure}
    \includegraphics[width=\figsize\textwidth]{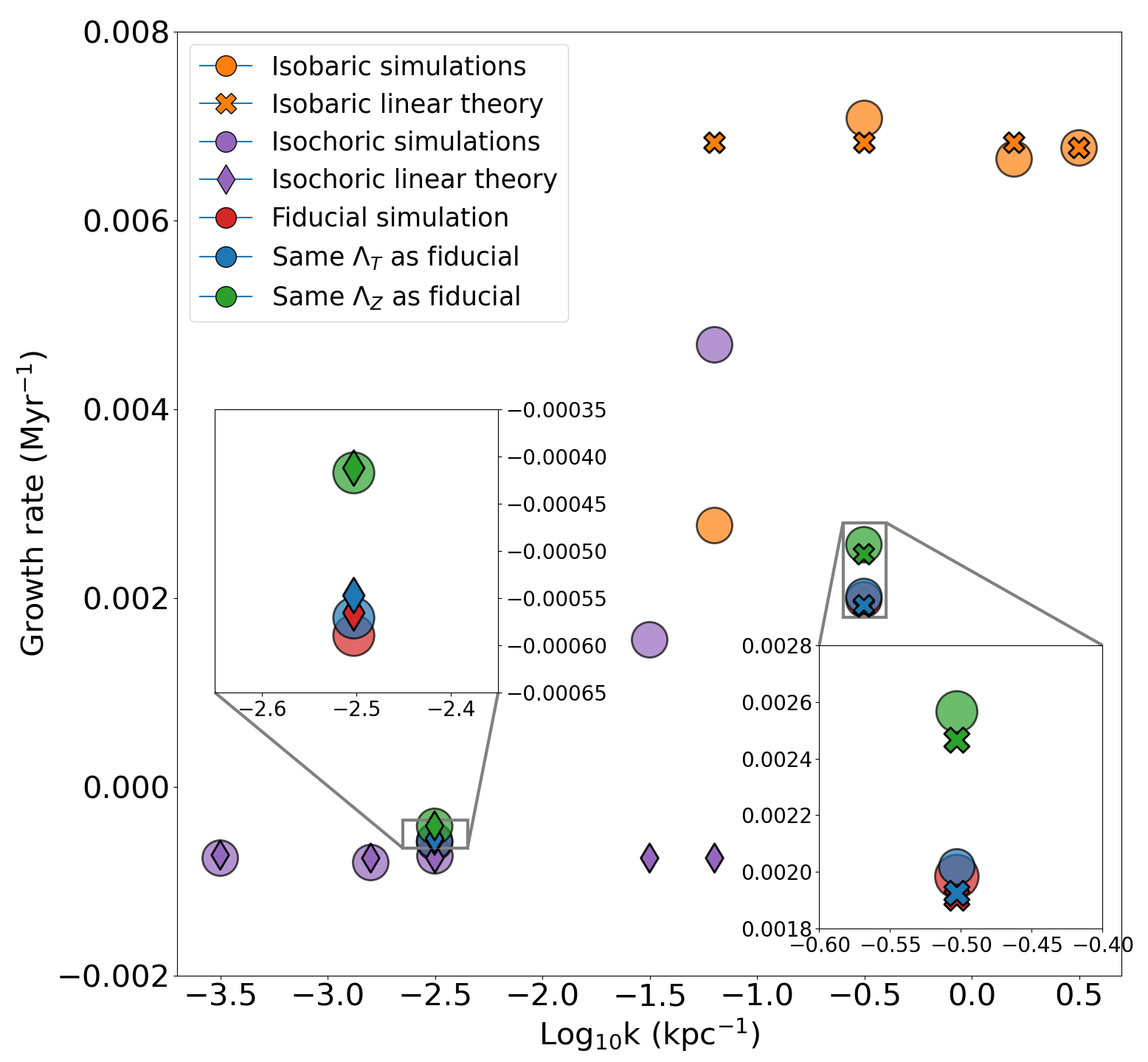}
    \centering
    \caption{The growth/damping rates for a range of wavenumbers ($k$) in both the isobaric and isochoric regimes, at T$\approx9\times10^6$ K. Orange circles and crosses show the simulation and the theoretical (asymptotic) growth rates, respectively, in the isobaric regime. Purple circles and diamonds show the simulation and theoretical (asymptotic) growth rates, respectively, in the isochoric regime. As expected, the simulations and linear theory agree in the isobaric/isochoric limits. {\it Inset}: Comparison of the growth rates calculated from simulations and linear theory for the three cases shown in Fig.~\ref{fig:lamcont} (with the same color code), in the isobaric and isochoric regimes. The growth/damping rates change with $\Lambda_T$ but not with $\Lambda_Z$, as expected from linear theory.}
    \label{fig:growth}
\end{figure}

Linear theory predicts the same growth rate (relative to $t_{\rm cool, 0}^{-1}$) for the pair with the same $\Lambda_T$ but different $\Lambda_Z$ both in isobaric and isochoric regimes, while for the other pair the growth rates should be different. This is indeed what is seen in the linear stage of the corresponding simulations (Fig.~\ref{fig:growth}). 

\begin{figure}
    \includegraphics[width=\figsize\textwidth]{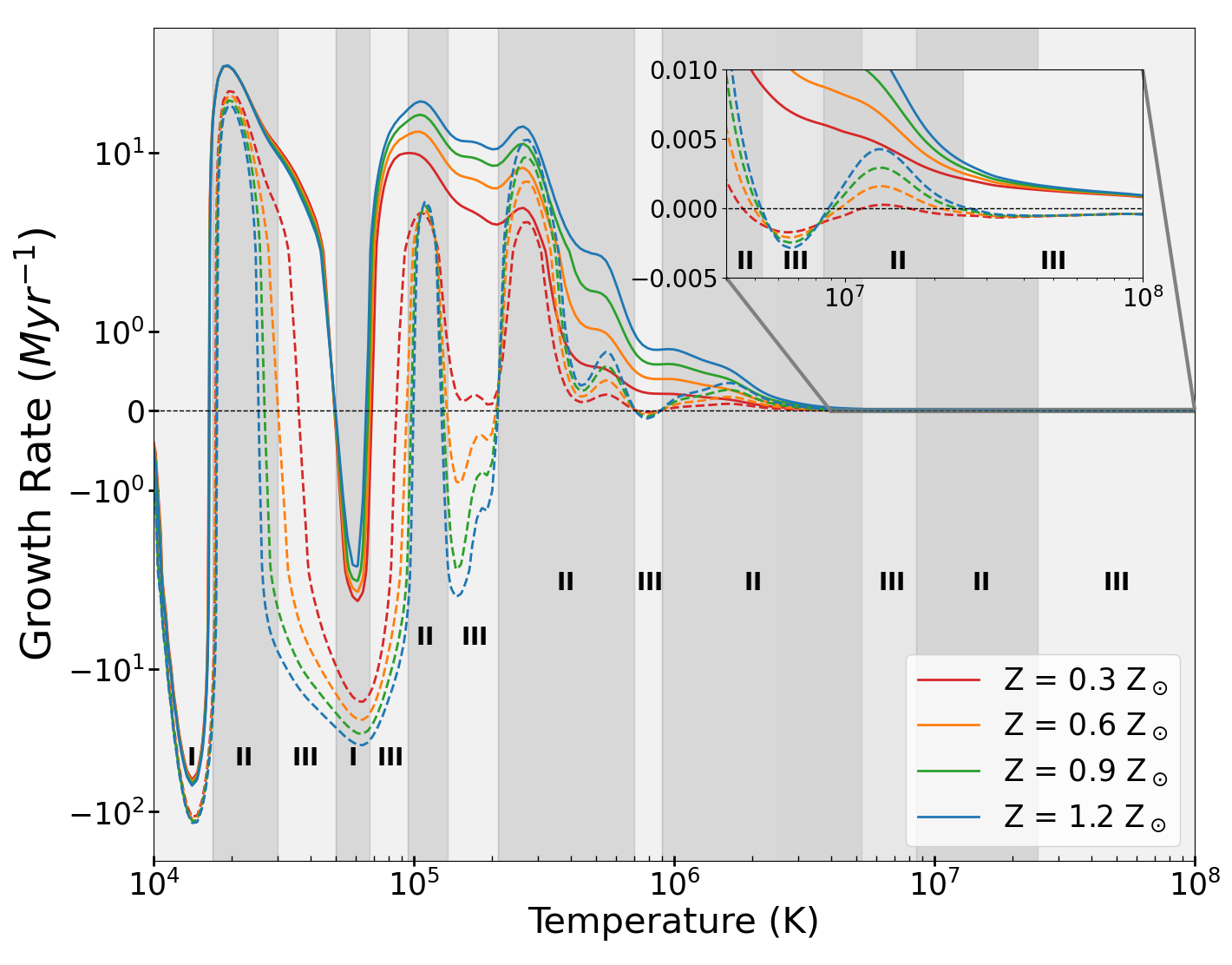}
    \centering
    \caption{Analytic asymptotic growth rates for isobaric (solid lines; Eq. \ref{eq:isobaric}) and isochoric (dashed lines; Eq. \ref{eq:isochoric}) modes as a function of temperature for different metallicities. The background density is 0.062$m_p$ g cm$^{-3}$. The inset shows the theoretical growth rates at different metallicities for a smaller temperature range. The shaded regions show different regimes depending on the temperature (these regions change slightly with metallicity; here shading corresponds to $0.6Z_\odot$). In region I, both the isobaric and isochoric modes are stable. In region II, both the isobaric and isochoric modes are unstable, and in region III the isobaric mode is unstable and the isochoric is stable.}
    \label{fig:growth_rate_theory}
\end{figure}
Fig.~\ref{fig:growth} shows the growth rate obtained from simulations and linear theory for a large range of $k$s. Note that the small scale fluctuations (large $k$s) are isobaric while the large scale fluctuations (small $k$s) are isochoric. 
To compare theory and simulations across $k$s, we consider $\ln(\delta\rho/\rho_0)$ versus time as in  Fig.~\ref{fig:rhoisobaric} for each $k$ and calculate (within linear regime) the slope using a linear fit. We repeat this process for all the simulations listed in Table \ref{tab:linear_sim} and plot in Fig.~\ref{fig:growth}. The simulation growth rates are shown with `circle' markers in different colors. The `cross' and `rhombus' markers show the corresponding values calculated from linear theory in the isochoric and isobaric regimes, respectively. The inset plots show the three cases with different $\Lambda_T$ and $\Lambda_Z$ shown in Fig.~\ref{fig:lamcont}. 
The green marker with the same $\Lambda_Z$ as the fiducial case (red circle in Fig.~\ref{fig:growth}) but a different $\Lambda_T$,  clearly has a higher (lower) growth (decay) rate. Thus the simulation results agree with the analytic expectation that the linear growth rate of thermal instability depends on $\Lambda_T$ but not on $\Lambda_Z$. 
The case of $k=\pi/(100~\rm kpc)$ and $\pi/(50~\rm kpc)$ are intermediate between the isobaric and isochoric regimes, and therefore the measured growth rate differs from the asymptotic limits. This trend of the linear growth rate versus wavenumber, going from the isochoric to isobaric regime, matches existing simulations (e.g., 
see Fig.~1 of \citealt{Piontek2004}).

\subsection{Nonlinear evolution}
\label{sec:nonlinear}

\begin{figure*}
    \includegraphics[width=18cm]{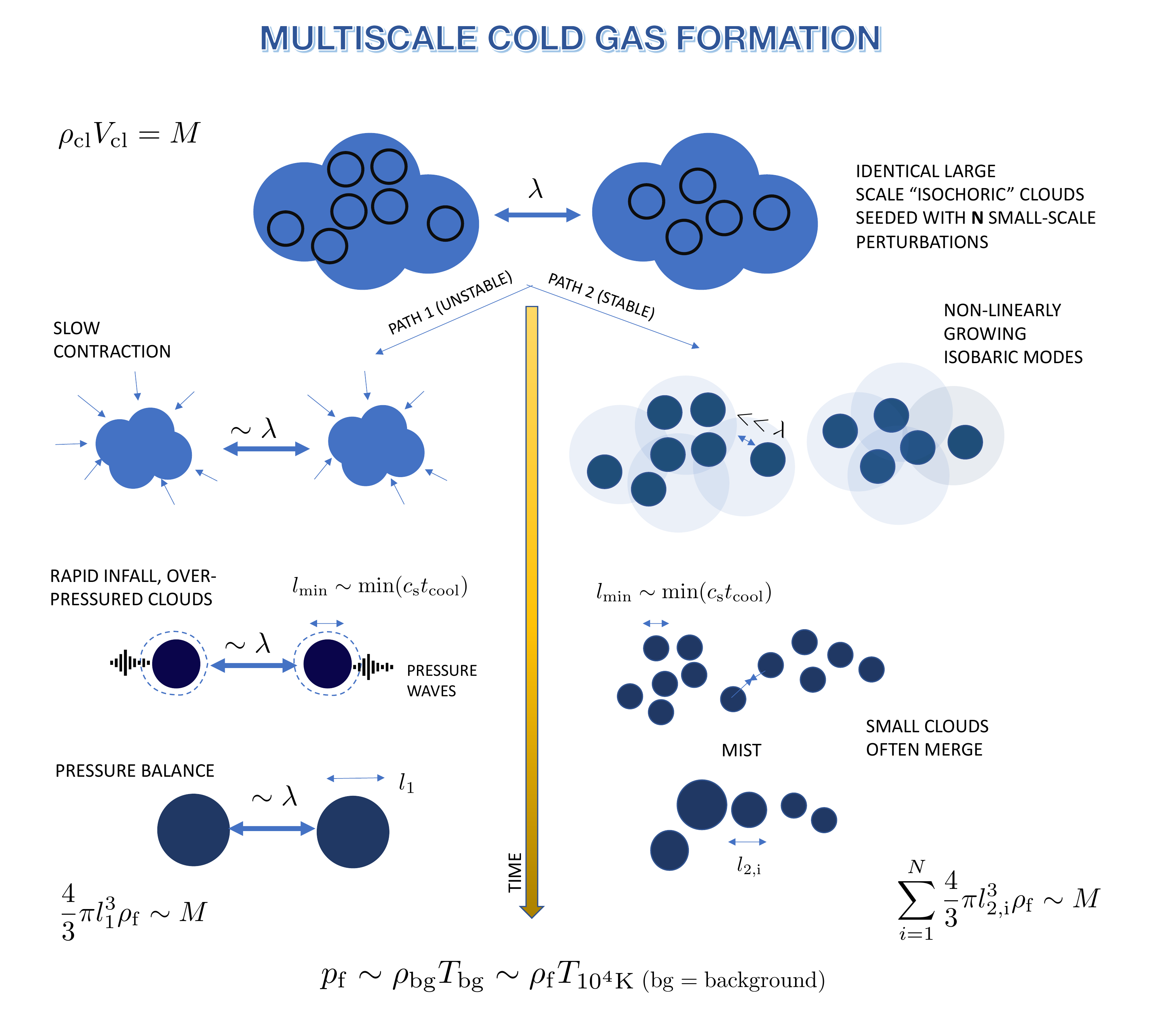}
    \centering
    \caption{Illustration of how isochoric clouds can evolve nonlinearly, depending on the stability of the linear isochoric mode. The linear stability of an isochoric cloud depends on the temperature (and metallicity to a smaller extent; Fig. \ref{fig:growth_rate_theory}). The large unstable clouds (the left pathway above) quickly lose pressure support (on roughly a cooling time) isochorically (at a constant density) and cool to the stable temperature. Eventually, after the much longer sound-crossing time across the cloud, the under-pressured clouds slowly  collapse to smaller volumes, erasing all the small scale isobaric growing modes. At the end of this collapse the cloud pressure overshoots, reflecting pressure (sound) waves as the gas achieves pressure equilibrium. The eventual cloud-size is small (determined by mass conservation of the collapsing volume) but the separation between clouds is of order the initial wavelength.  
    On the contrary, the stable large scale clouds (the right pathway above) decay and become sites for the growth of "misty" small scale isobaric modes clustered closely (with cold gas again at the stable phase but in pressure balance). While the cloudlets are small ($\sim {\rm min}[c_s t_{\rm cool}]$) in the transient state, most of them merge at late times (see Fig. \ref{fig:cloud_size_t}). 
    }
    \label{fig:multiscale_diag}
\end{figure*}

\begin{figure}
    \includegraphics[width=\figsize\textwidth]{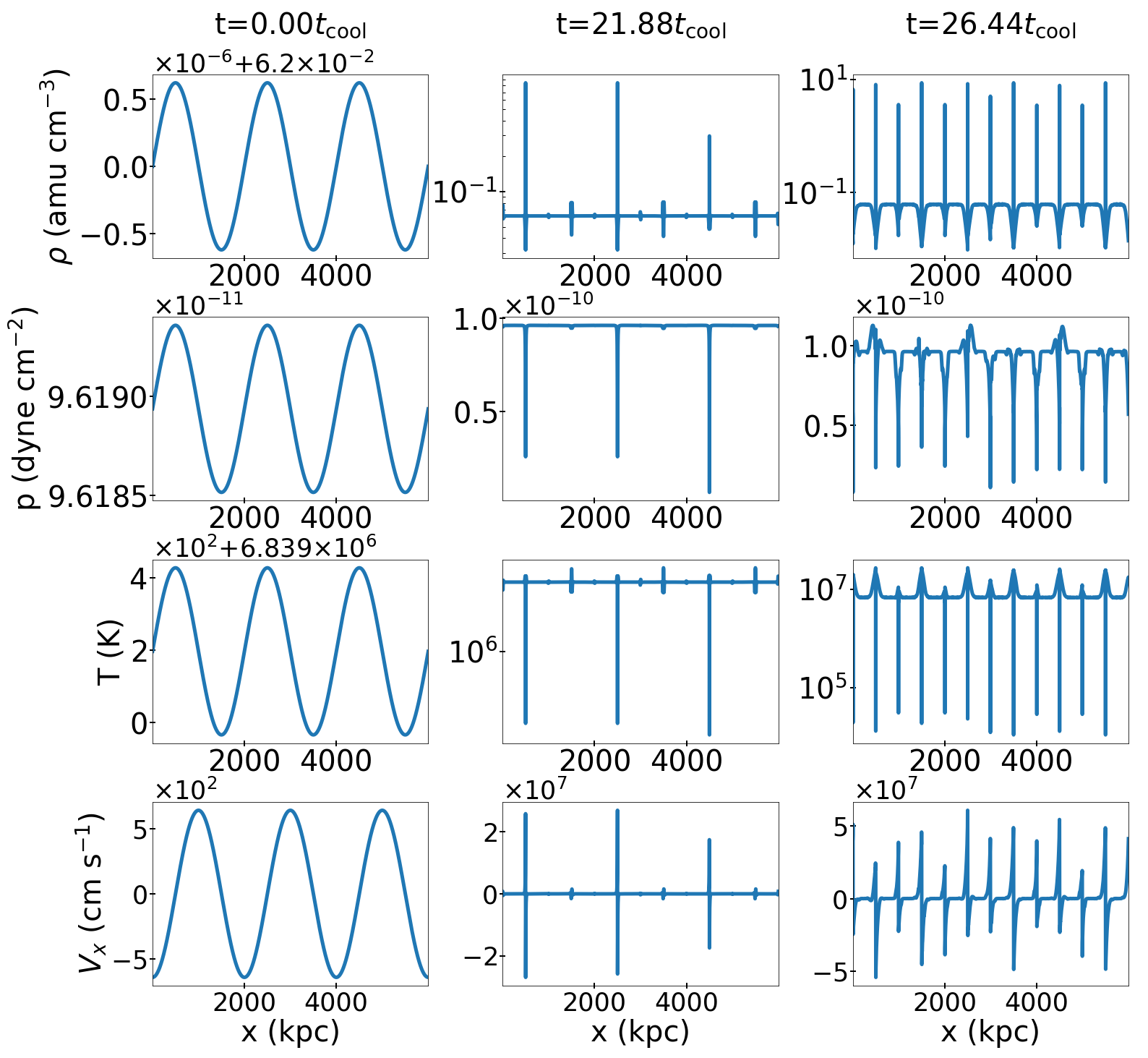}
    \centering
    \caption{Density, pressure, temperature, and velocity profiles for the stable isochoric run with periodic boundary conditions (NIC\_ST\_P) at different times showing the initial condition, nonlinear growth, and the (quasi)steady state. 
    Grid-scale isobaric modes grow and dominate in the nonlinear state, resulting in dense, cold gas separated by less than the wavelength of the initial isochoric perturbation.
    }
    \label{fig:nonlin_ic_st_periodic}
\end{figure}

\begin{figure}
    \includegraphics[width=\figsize\textwidth]{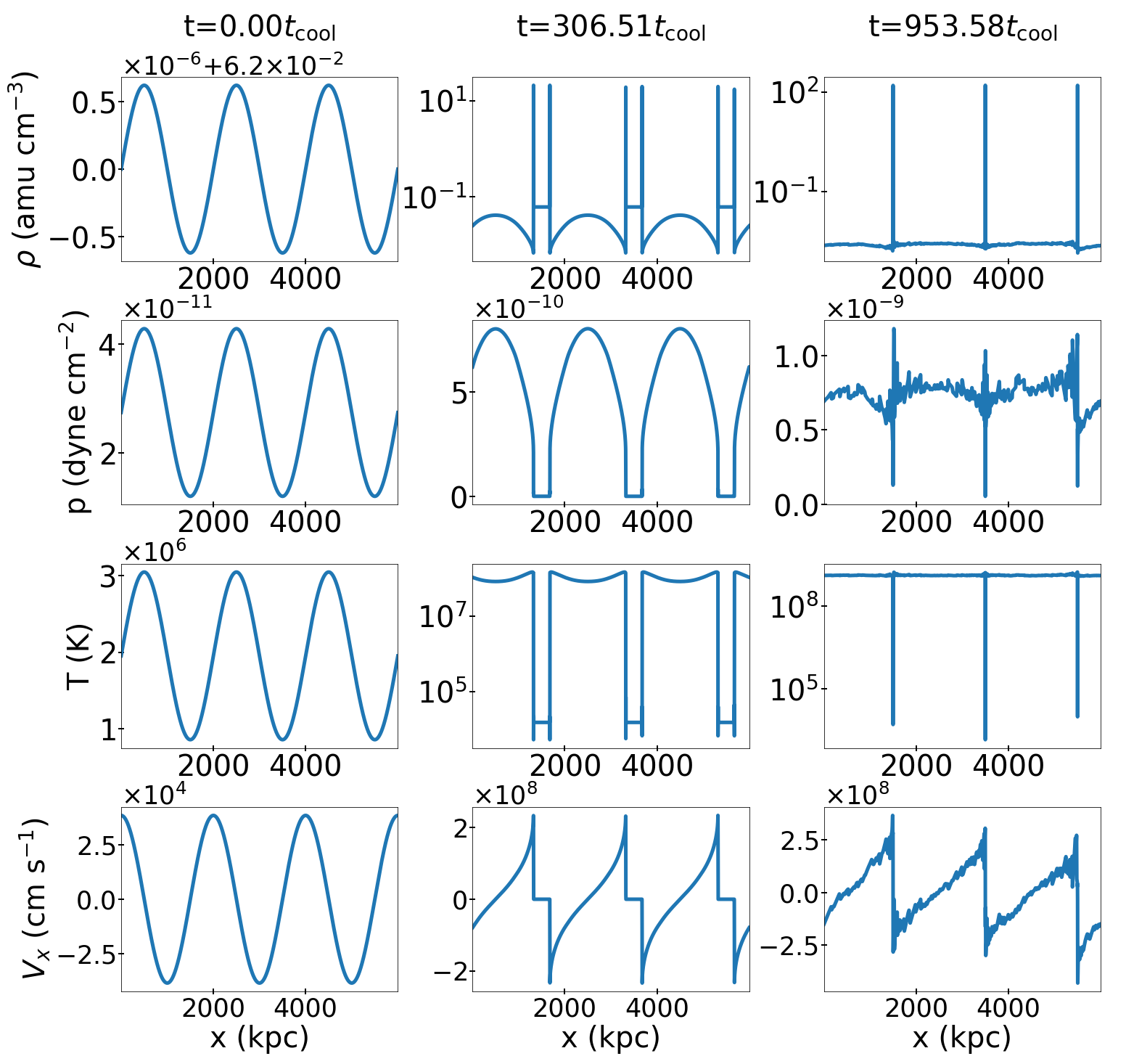}
    \centering
    \caption{Density, pressure, temperature, and velocity profiles for the unstable isochoric run with periodic boundary conditions (NIC\_UST\_P) at different times showing the initial condition, nonlinear growth and the (quasi)steady state. 
    }
    \label{fig:nonlin_ic_ust_periodic}
\end{figure}

Now that we have verified linear theory with simulations, we proceed to study the nonlinear evolution of thermal instability. We find that the linear response also plays a crucial role in the nonlinear evolution of the condensation mode. Namely, the nonlinear growth of perturbations at small (isobaric) scales, within a large scale perturbation, is only possible if the isochoric mode is linearly stable. For an unstable isochoric mode we do not observe growth of multiphase gas at small scales. 

Since linear behavior is important to understand the nonlinear evolution of thermal instability, we start this section with Fig.~\ref{fig:growth_rate_theory}, which shows the  theoretical (asymptotic) growth rates for the isobaric and isochoric modes as a function of temperature with different metallicities. Depending on the temperature and metallicity, we define three regimes of thermal instability: stable isochoric and isobaric modes (region I); unstable isochoric and isobaric modes (region II, relevant for the CGM of the Milky way with $T\sim 10^6$ K); and stable isochoric and unstable isobaric mode (region III, relevant for the ICM with $T>{\rm few}\times 10^7$ K). We investigate the two regions with unstable isobaric modes in more detail. Figure \ref{fig:multiscale_diag} 
illustrates the different nonlinear evolution of the stable and unstable isochoric clouds, leading to the formation of 
cold gas in the two scenarios. These scenarios presented in the cartoon are corroborated by the nonlinear simulations (see Table~\ref{tab:nonlinear_sim}) presented next. 


\subsubsection{Periodic runs}
To check for any difference in the nonlinear evolution of the stable (NIC\_ST) and unstable (NIC\_UST) isochoric runs, we run them without a metallicity gradient and with periodic boundary conditions (NIC\_ST\_P and NIC\_UST\_P, respectively). Note that in the absence of metallicity and H abundance gradients, there should be no  perturbation in metallicity and H abundance (Eqs. \ref{eq:dye}, \ref{eq:dye_a}). Moreover, for a passive scalar metallicity we do not expect the metallicity profile to evolve.

Figures \ref{fig:nonlin_ic_st_periodic} and \ref{fig:nonlin_ic_ust_periodic} show the density, pressure, temperature, and velocity profiles for the stable and unstable isochoric runs (NIC\_ST\_P, NIC\_UST\_P) with periodic boundary conditions at different times. It is evident that the nonlinear evolution for these two is very different. For the stable isochoric run, the large scale isochoric mode decays but isobaric disturbances start growing at the grid scale close to the density fluctuation extrema and zeros. The small scale density peaks in the nonlinear state (right panels of Fig. \ref{fig:nonlin_ic_st_periodic}) are separated by less than the wavelength of the initial isochoric mode. For the unstable isochoric mode, on the other hand, the large scale mode becomes nonlinear without much growth of isobaric modes at small scales. The supersonic gas falling on to the cold isochoric regions forms a radiative shock (e.g., see Chapter 16 in \citealt{Shu1992}), with a large density jump at the boundary of the isochoric cold cloud. At late times (right panels of Fig. \ref{fig:nonlin_ic_ust_periodic}) the cold, under-pressurized regions collapse, and pressure equilibrium is established as the radiative shocks merge. Notice the oscillations in the ambient hot gas launched due to cloud-collapse. The final dense peaks in this case are separated by the wavelength of the initial isochoric mode.


When the overdense regions are isobaric, the high (low) temperatures correspond to low (high) density regions. However, this is not true for the unstable isochoric modes till they shrink and become isobaric at late times (right panels of Fig. \ref{fig:nonlin_ic_ust_periodic}). For the unstable isochoric modes (NIC\_UST\_P), we notice the formation of large cold regions ($\sim 100$s kpc) which shrink in size with time (Fig. \ref{fig:nonlin_ic_ust_periodic}). For all runs, the velocity becomes discontinuous and changes direction in the nonlinear state, indicating the accretion of matter on to the dense regions. But for the unstable isochoric run NIC\_UST\_P, at the onset of nonlinearity (bottom-middle panel), we see a region of zero velocity sandwiched by infalling gas with opposite signs. The static cold sandwiched regions compress and later evolve into a discontinuity similar to the nonlinear isobaric structures (e.g., see the isobaric run NIB\_OF in Fig. \ref{fig:nonlin_ib}). This is different from the stable isochoric run NIC\_ST\_P, where the velocity switches sign abruptly across the thin cold layer even at early times (middle column of Fig. \ref{fig:nonlin_ic_st_periodic}). 

%
%
%
\subsubsection{Runs with metallicity gradient}
To further investigate the nonlinear evolution of the stable and unstable isochoric modes, we add metallicity gradient and metallicity perturbations with outflow boundary conditions. 
\begin{figure}
    \includegraphics[width=\figsize\textwidth]{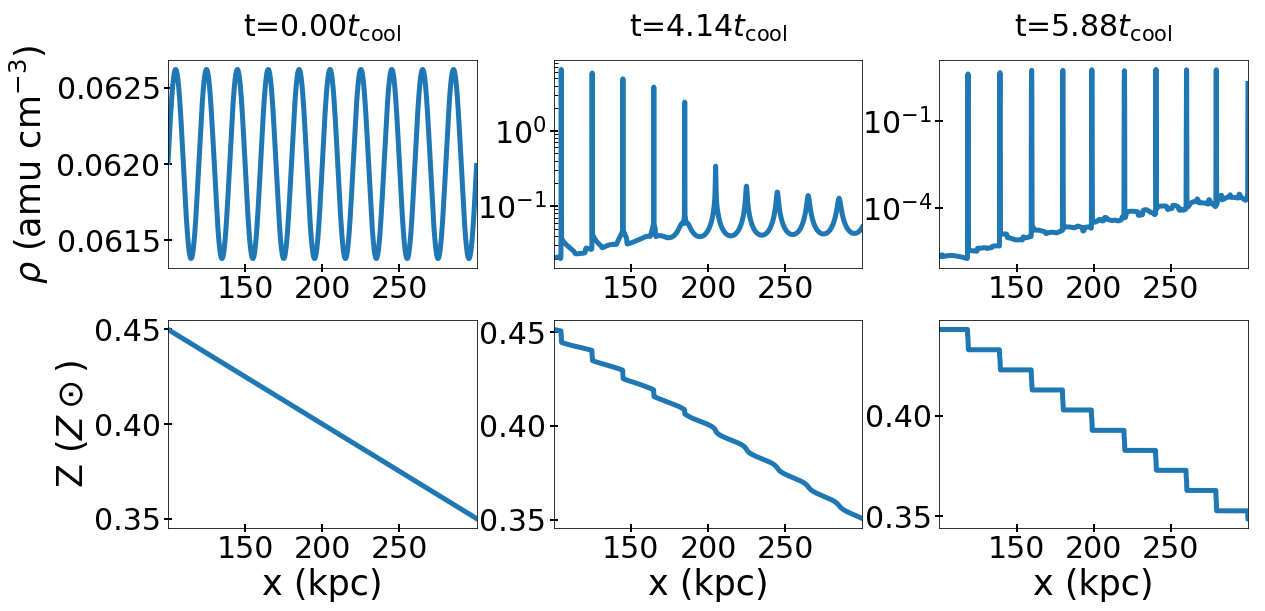}
    \centering
    \caption{Density and metallicity profiles for the isobaric modes with a background metallicity gradient and outflow boundary conditions (NIB\_OF) at different times. (We do not show the buffer zones close to  boundaries where cooling and heating are turned off). 
    The perturbations become nonlinear first at smaller $x$ where
    the metallicity is largest and the cooling time is shortest. The perturbations grow "in place" and there is no fragmentation. 
    Notice the stair-case pattern in the metallicity at late times, which can be explained from the advection of the background metallicity by flows on to dense regions.
    }
    \label{fig:nonlin_ib}
\end{figure}
\begin{figure}
    \includegraphics[width=\figsize\textwidth]{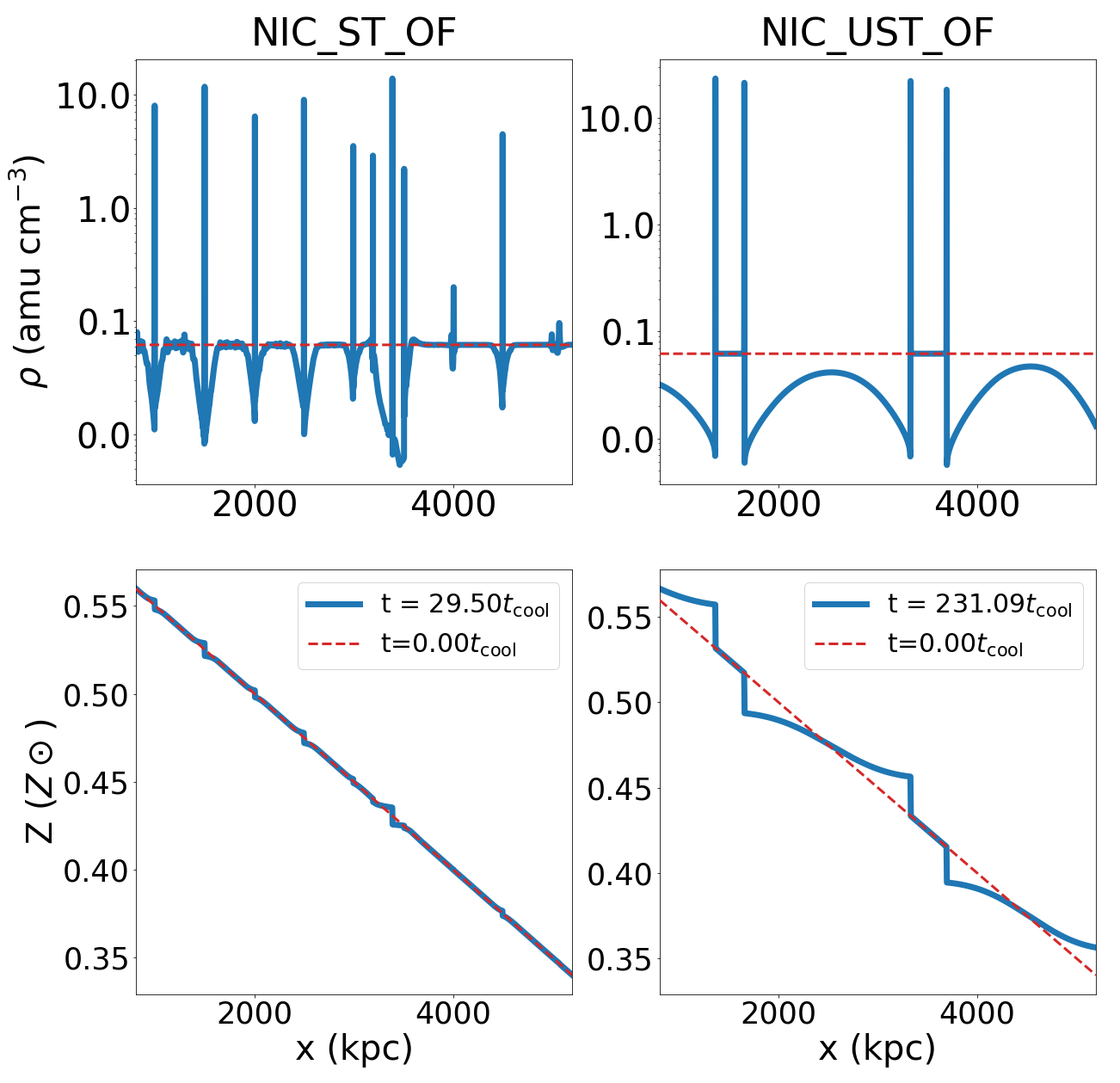}
    \centering
    \caption{Density and metallicity profiles for the stable and unstable isochoric run with a background metallicity gradient and outflow boundary conditions (NIC\_ST\_OF [left panels], NIC\_UST\_OF [right panels]) at different times. The initial profiles are shown with red dashed lines. (We do not show the buffer zones where we turn off cooling and heating). 
    Similar to the stable isochoric mode with periodic boundary conditions (in Fig. \ref{fig:nonlin_ic_st_periodic}), we see the growth of isobaric perturbations at the grid scale and the separation between the dense clouds less than the wavelength of the initial isochoric mode. The "steps" in the metallicity profile at late times are smaller because this run, having isochoric modes, is not as deep into the nonlinear state as compared to the the bottom-right panel in Fig. \ref{fig:nonlin_ib} (compare the density profiles in the two cases). Unlike the stable isochoric mode, nonlinearly we do not see growth of multiphase gas at small scales in the unstable isochoric mode. Again we see the staircase pattern in metallicity at late times. Notice that the metallicity is unchanged in the cold under-pressured regions which have not collapsed yet.
    }
    \label{fig:nonlin_ic_outflow}
\end{figure}

Fig.~\ref{fig:growth_rate_theory} shows that for a given metallicity, the growth rate of isobaric modes is always higher compared to the isochoric modes. In the run with an initially isobaric mode with a background metallicity gradient (NIB\_OF; Fig.~\ref{fig:nonlin_ib}), the overdense regions grow nonlinearly and eventually form dense peaks with densities 3 orders of magnitude larger than the surrounding diffuse medium. The growth of nonlinear structure is the fastest in the highest metallicity regions because the cooling time is the shortest there (and hence the growth rate of thermal instability is the fastest). Nonlinearly, the surrounding gas cools and accretes on to the dense seeds. Also notice the stair-case structure of metallicity in the extreme nonlinear state (bottom-right panel). These can be understood from the passive scalar equation satisfied by metallicity (Eq. \ref{eq:initial3}) and by the tendency of matter to accrete on to dense seeds in the nonlinear state.

For the stable isochoric mode with an initial metallicity gradient (NIC\_ST\_OF; left panels of Fig.~\ref{fig:nonlin_ic_outflow}), the initially isochoric mode decays, but the unstable isobaric peaks, seeded by truncation errors, start to grow at small scales. 
In the nonlinear state, we see the formation of numerous other isobaric peaks (as in the periodic stable isochoric modes in Fig. \ref{fig:nonlin_ic_st_periodic}). 
For the unstable isochoric mode with a background metallicity gradient (NIC\_UST\_OF;  right panels of Fig.~\ref{fig:nonlin_ic_outflow}), 
we do not see the formation of 
isobaric peaks at the beginning of the nonlinear stage. The isobaric peaks formed in the linear state (not shown) come close and merge to give rise to a single  dense peak for each initially isochoric overdense region. 

The metallicity profiles show a stair-case pattern 
in the nonlinear steady state for all three cases (isobaric, and stable and unstable isochoric with metallicity gradient). In the isobaric and unstable isochoric modes with a metallicity gradient (NIB\_OF and NIC\_UST\_OF), during the nonlinear growth stage, the slope of the metallicity profile decreases between discontinuities (bottom panels of Figs. \ref{fig:nonlin_ib}, \ref{fig:nonlin_ic_outflow}). This is due to the infall of matter into overdense peaks. In Eq.~\ref{eq:initial3}, the gradient of metallicity has a negative sign. Thus, the points where velocity is positive (negative) , $\partial Z/\partial t$ is positive (negative). From the initial metallicity profile and the velocity profiles from the periodic runs, we deduce that this should result in flattening of the metallicity profile between the overdense peaks. In the stable isochoric run (NIC\_ST\_OF), the flattening of the metallicity profile between density discontinuities is less pronounced (bottom panels of Fig. \ref{fig:nonlin_ic_outflow}) 
as profiles are shown at a much earlier time in comparison to the unstable isochoric run with a metallicity gradient (NIC\_UST\_OF). As in the periodic simulations (Figs. \ref{fig:nonlin_ic_st_periodic} \& \ref{fig:nonlin_ic_ust_periodic}), the metallicity profiles are discontinuous across the density peaks for the stable isochoric run and they are unmodified within the cold collapsing peaks for the unstable isochoric run.  


\subsubsection{Evolution of cold peaks}
\begin{figure*}
    \includegraphics[width=18cm]{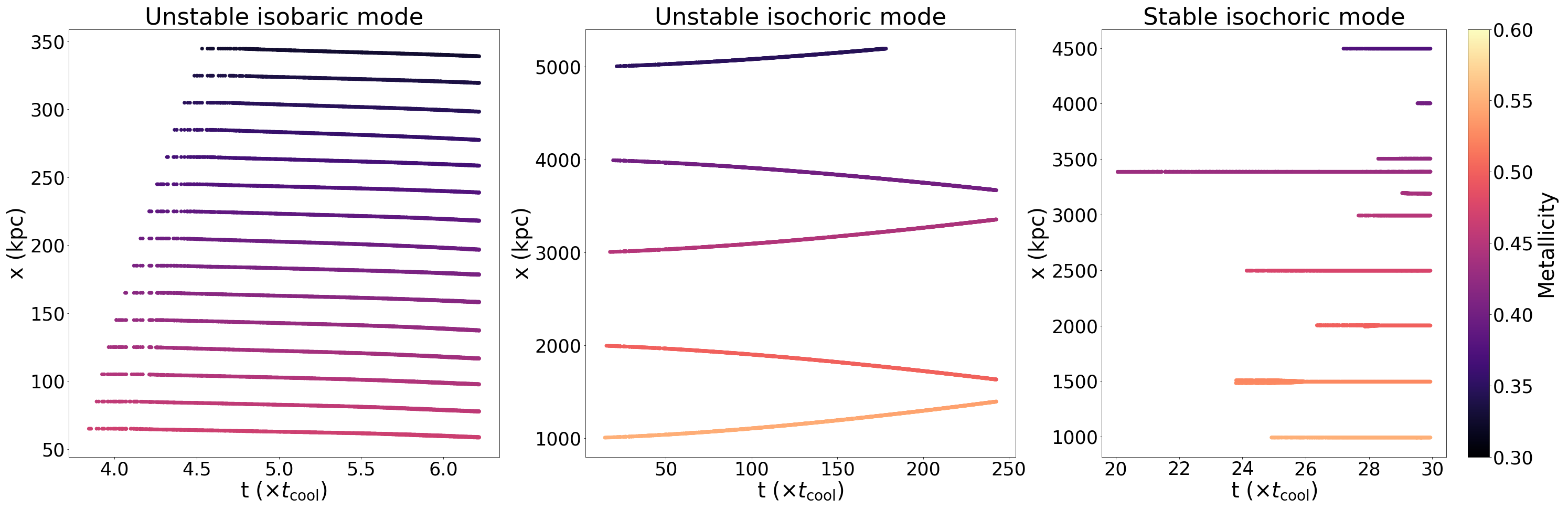}
    \centering
    \caption{Position of grid points with cold gas ($<10^{4.1}$ K) as a function of time for various runs: isobaric (NIB\_OF; Fig. \ref{fig:nonlin_ib}), unstable isochoric (NIC\_UST\_OF; Fig. \ref{fig:nonlin_ic_outflow}), and  stable isochoric (NIC\_ST\_OF; Fig. \ref{fig:nonlin_ic_outflow}). Color indicates the metallicity of the cold gas. This figure  shows the difference in the evolution of cold gas in the three cases. The growth of small-scale cold gas for stable isochoric mode and the absence of small scale cold gas in the unstable isochoric mode is easily seen. Nonlinearly, the dense peaks can merge due to local pressure gradients, especially the cold, high pressure peaks flanking the low temperature, under-pressured region of the unstable isochoric run (middle panel).}
    \label{fig:cold_mass}
\end{figure*}

Figure \ref{fig:cold_mass} shows the evolution of cold ($<10^{4.1}~\rm K$) gas with time for the isobaric (NIB\_OF), and unstable (NIC\_UST\_OF) and stable (NIC\_ST\_OF) isochoric runs. Color indicates the metallicity of the cold gas.  Figure~\ref{fig:cold_mass} (left panel) shows that cold gas condenses at the positions of the initial overdense peaks, which then grow nonlinearly. Generally, the cold gas arises the earliest in the highest metallicity region (this is not always the case, e.g., in the right panel the metallicity of the earliest cold gas is not the highest). The initial density and temperature are also important factors, in addition to metallicity.
For the unstable isochoric mode, the dense peaks come close together with time as the central cool region is under-pressured and collapses due to the external pressure (see the right panels of Fig. \ref{fig:nonlin_ic_outflow}). In contrast, the stable isochoric mode shows the growth of unstable isobaric mode and cold gas at smaller scales than the wavelength of the initial mode (see the left panels of Fig. \ref{fig:nonlin_ic_outflow}). On a much longer timescale, we expect the dense peaks to merge together because the regions surrounding them have a low pressure due to radiative losses (e.g., see the right panel of Fig. 2 in \citealt{Sharma2010}; 
see also, \citealt{Waters2019b}).

\subsection{Isochoric stability as the condition for small scale cold gas}
\label{sec:shattering}
We observe different cold gas evolution for the stable and unstable isochoric modes. For stable isochoric modes, we observe the growth of small-scale isobaric thermal instability and cold gas with separation smaller than the wavelength of the isochoric perturbation. This is different from
the shattering envisaged in \citet{McCourt2018} because we do not see the nonlinear fragmentation of the densest isochoric regions with short cooling times. For the unstable isochoric modes, we do not observe any growth of small-scale (isobaric) perturbations in the nonlinear state but the eventual collapse of the cold isochoric regions as they become isobaric and launch sound waves in the process (as suggested by \citealt{Waters2019}). We ran two simulations with identical initial conditions but different power law cooling functions (not discussed in detail) such that the isochoric mode is stable or unstable for all temperatures. We observe the same correlation between the stability of the isochoric mode and the nonlinear cold gas evolution.



\subsubsection{Runs with a different background temperature}
Fig.~\ref{fig:growth_rate_theory} shows that there are regions near $10^5~\rm K$ where, depending on the metallicity, the isochoric mode can be stable or unstable. This dependence on metallicity is due to the  dependence of the growth rate on metallicity through $\Lambda_T$ (Eq. \ref{eq:isochoric}).

\begin{figure}
    \includegraphics[width=\figsize\textwidth]{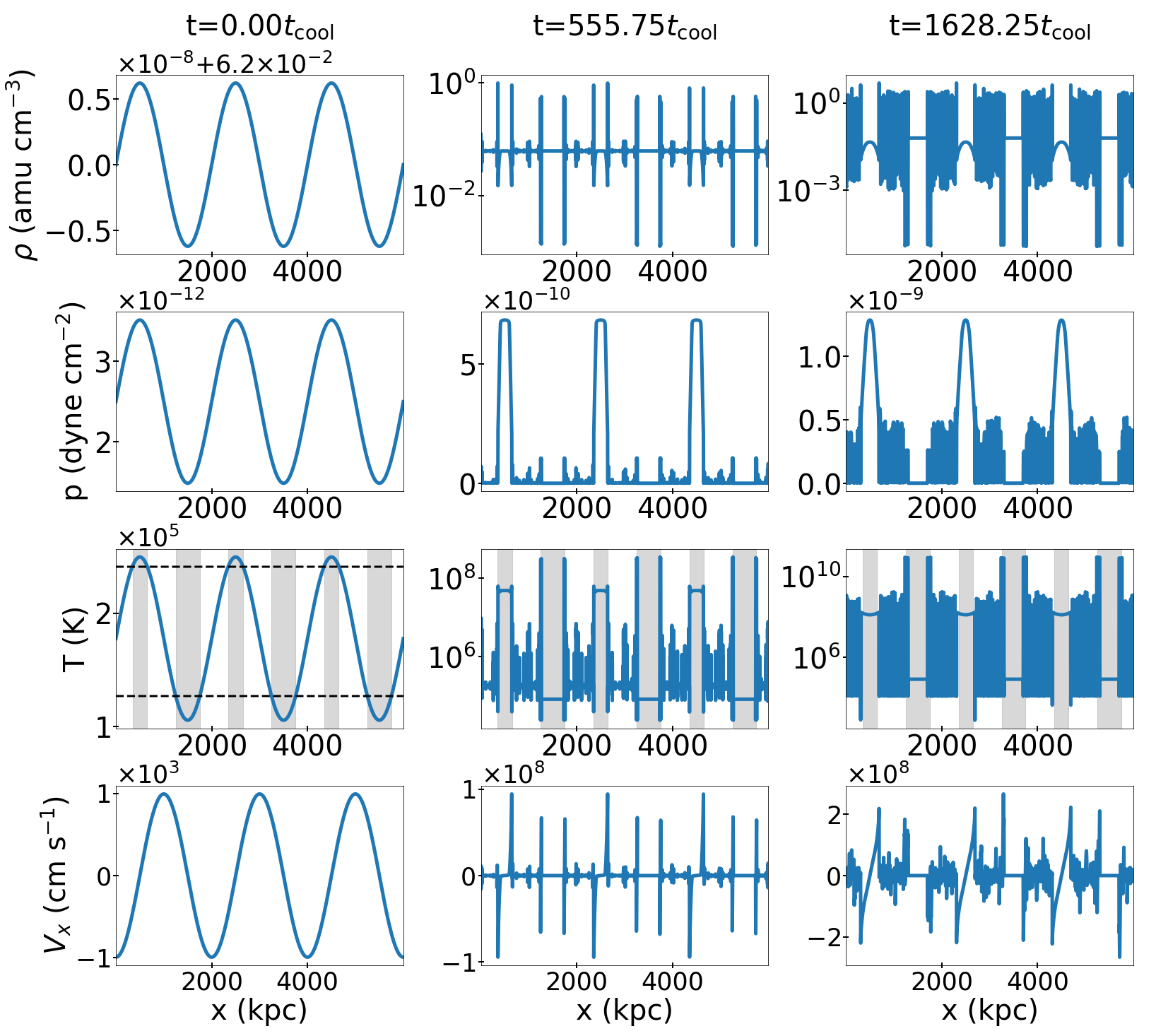}
    \centering
    \caption{Density, pressure, temperature, and velocity profiles for the periodic low-temperature run at different times with both stable and unstable isochoric temperatures (NIC\_P\_LOWT). The locations with a temperature range corresponding to the stable isochoric regime ($1.2 \times 10^5~{\rm K}<T<2.3 \times 10^5$ K; see Fig. \ref{fig:growth_rate_theory}) show the growth of small scale isobaric perturbations. Thus the linear properties of the isochoric mode also determine the nonlinear evolution of the cold gas.  Note that the shaded unstable isochoric temperature range does not undergo condensation at small scales.
    }
    \label{fig:nonlin_ic_stable_lowT}
\end{figure}

To test whether the stability or instability of the isochoric mode determines the appearance of multiphase gas in the nonlinear state, we run a periodic simulation of a stable isochoric eigenmode with the initial temperature ranging from $\approx 10^5$ K to $\approx 2.5 \times10^5 \rm K$. Fig.~\ref{fig:nonlin_ic_stable_lowT} shows the profiles from this simulation. According to Fig.~\ref{fig:growth_rate_theory}, the isochoric mode is stable between $1.2-2.3 \times 10^5$ K. 
The regions with maximum and minimum temperature, being isochorically unstable, do not show the growth of small scale isobaric modes. While the intermediate temperature regions, sandwiched between the temperature extrema, being isochorically stable, show the growth of isobaric perturbations at small scales.
Thus we see that nonlinear evolution, even locally, depends on whether the isochoric mode is linearly stable or unstable.


\begin{figure*}
    \includegraphics[width=18cm]{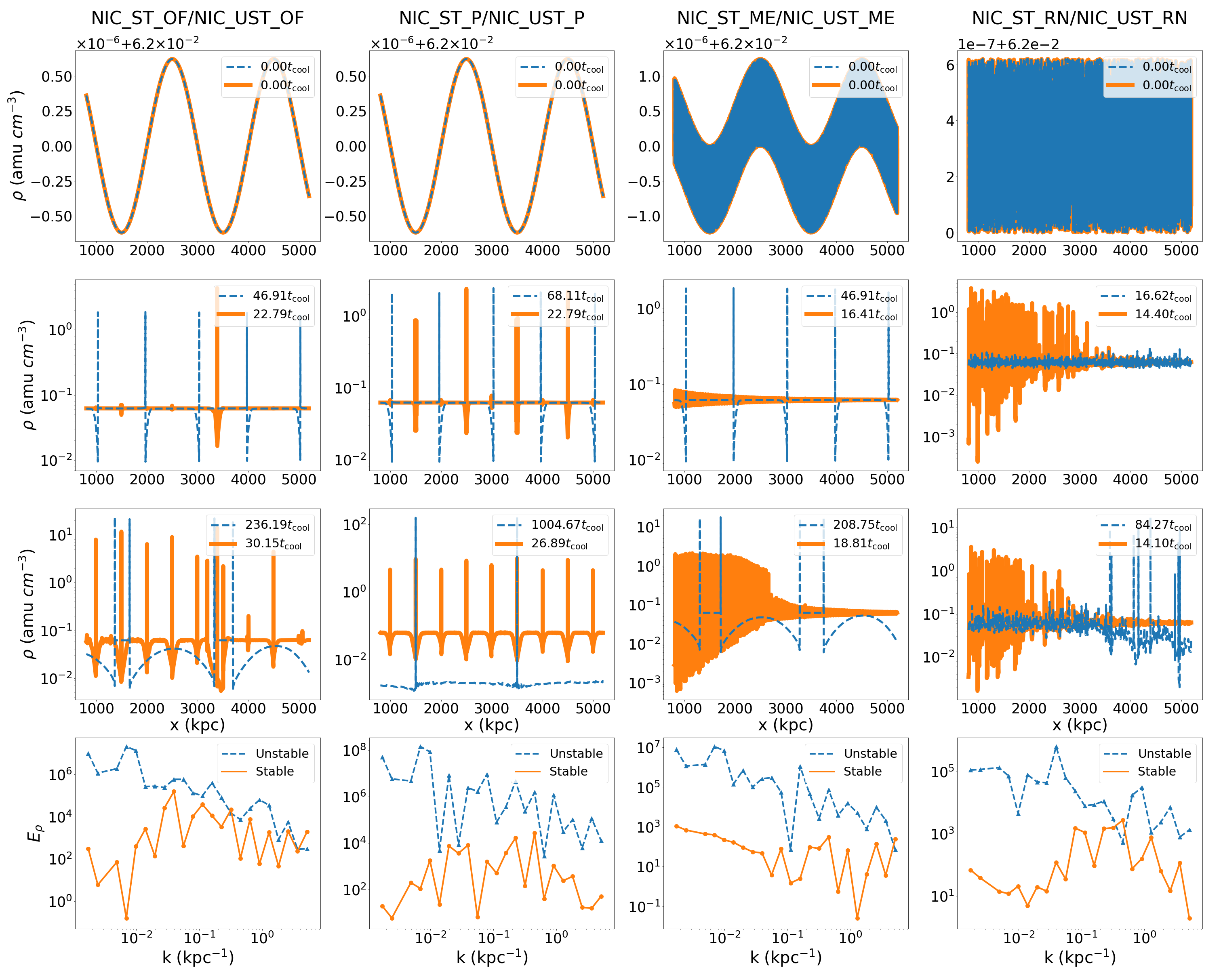}
    \centering
    \caption{The top three rows show the density profiles at various times for different runs with unstable (blue dashed lines) and stable isochoric modes (orange solid lines; see Table \ref{tab:nonlinear_sim}). While the runs shown in the second column have a constant background metallicity, the others have a background metallicity gradient. The last row shows the power spectrum of the density profiles shown in the third row. These plots show a significant difference between the runs with stable and unstable isochoric modes. Namely, that the multiphase gas predominantly grows at small-scales if the isochoric mode is stable. The power spectrum plots also corroborate this, which show much larger power at small $k$s if the isochoric mode is unstable. In the two right columns, with the stable isochoric modes, the low metallicity regions at large $x$ have not become as nonlinear as the high metallicity regions at small $x$. This gives rise to a small increase in the density power spectrum at small $k$s for these runs.}
    \label{fig:nonlin_ic_fourier}
\end{figure*}

\subsubsection{Mixed eigenmodes}
\label{sec:mixed_eig}
\begin{figure}
    \includegraphics[width=\figsize\textwidth]{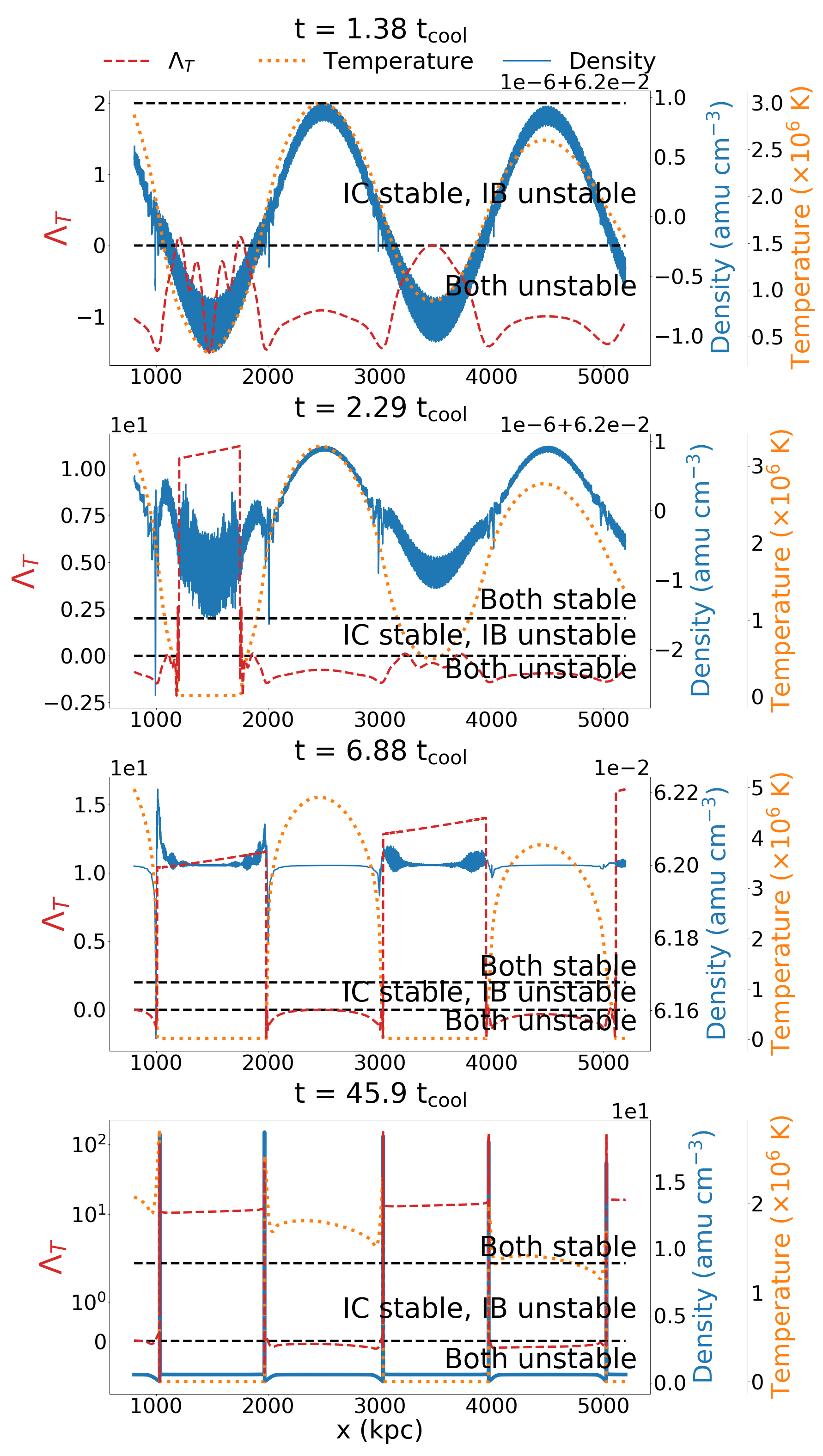}
    \centering
    \caption{The profiles of density, temperature, and the logarithmic temperature derivative of the cooling function ($\Lambda_T \equiv \partial \ln \Lambda/\partial \ln T$) for the unstable isochoric mixed-eigenmode run (NIC\_UST\_ME) at different times. We label the different regions along y-axis according to the stability of the isobaric and isochoric modes. The stability (instability) of
    the isochoric (isobaric) mode explains the nonlinear growth of the small-scale isobaric modes only at specific regions in space where $2>\Lambda_T>0$.}
    \label{fig:lamT_rho}
\end{figure}



We run additional simulations with mixed eigenmodes and random noise initial conditions (two right panels in Fig. \ref{fig:nonlin_ic_fourier}), to test the influence of the linear stability of the isochoric mode on the nonlinear evolution of multiphase gas. We also calculate the density power spectrum (bottom panels of Fig. \ref{fig:nonlin_ic_fourier}) for all the different cases to quantify the structure of the multiphase gas in the nonlinear state. This figure repeats in columns 1 \& 2 profiles from the stable and unstable isochoric runs with and without metallicity gradient for comparison (profiles of these runs are shown in Figs. \ref{fig:nonlin_ic_st_periodic}, \ref{fig:nonlin_ic_ust_periodic}, \ref{fig:nonlin_ic_outflow}).

Two runs are initialized with stable and unstable isochoric large-scale ($k=2\pi$/[2 Mpc]) modes, respectively. In addition to these, we also initialize small scale ($k=2\pi$/[2 kpc]) unstable isobaric modes.
These two runs use different temperatures (see Table \ref{tab:nonlinear_sim}) to ensure that the isochoric mode is stable for one and unstable for the other (NIC\_ST\_ME and NIC\_UST\_ME, respectively). We initialize a sum of two eigenmodes from linear theory with equal amplitude of the density perturbation, added with a zero phase difference. Figure \ref{fig:nonlin_ic_fourier} shows a comparison of density evolution of various runs with stable and unstable isochoric modes. The nonlinear evolution of the mixed-mode run with an unstable isochoric mode (NIC\_UST\_ME; blue-dashed lines in third column) is similar to the single unstable isochoric mode run (NIC\_UST\_OF; blue-dashed lines in the first column). Namely, there is no growth of small-scale isobaric perturbations in the nonlinear state. In contrast, for runs with stable isochoric modes, multiphase gas is observed nonlinearly at small scales. This is true for all cases shown in Figure \ref{fig:nonlin_ic_fourier}. Thus, the linear behavior of the isochoric mode seems to determine whether we see condensation at small scales nonlinearly.

Figure \ref{fig:lamT_rho} shows the profiles of density, temperature and $\Lambda_T$ for the mixed eigenmode run with an unstable isochoric mode (shown by blue-dashed lines in the third column of Fig. \ref{fig:nonlin_ic_fourier}). The snapshots show the early and nonlinear evolution of these quantities. At early times $\Lambda_T<0$ and both the isobaric and isochoric modes grow. Nonlinearly, one can see that small scales perturbations grow where the isobaric mode is linearly unstable and the isochoric mode is stable. At late times, the small-scale fluctuations are wiped out where the isochoric mode is unstable. In regions where both the isochoric and isobaric modes are stable, the small scale fluctuations are slowly damped away (see the third panel of Fig. \ref{fig:lamT_rho}). Thus the late-time evolution of the single and multi-mode perturbations with unstable isochoric modes (blue dashed lines) first and third columns of Fig. \ref{fig:nonlin_ic_fourier}) is very similar. Namely, multiphase condensation at small scales is suppressed.


\subsubsection{Runs with random noise perturbations}
\label{sec:random_eig}
To further test the robustness of our criterion for the growth of isobaric modes at small scales,
we initialize stable and unstable isochoric background with small-amplitude ($\delta \rho/\rho_0=10^{-5}$) random density perturbations (drawn from uniform distribution in the range $\rho_0 - \delta \rho$ and $\rho_0 + \delta \rho$) at every grid point (see the last column in Fig. \ref{fig:nonlin_ic_fourier}). The other variables (velocity, pressure) are initialized with an amplitude corresponding to an isochoric eigenmode with $k = 2\pi/[2 \rm Mpc]$. For these NIC\_ST\_RN and NIC\_UST\_RN runs we observe a very different evolution in the nonlinear state. In NIC\_ST\_RN, similar to the mixed eigenmode runs, we see that the isobaric overdense peaks grow "in place". In NIC\_UST\_RN, we see that not all the overdense peaks grow and the peaks are more distant from each other and have a larger amplitude in comparison to NIC\_ST\_RN. To put it another way, we observe these peaks repeating over a larger length scale in NIC\_UST\_RN than in NIC\_ST\_RN, as seen in other unstable isochoric simulations.

The various runs in Fig. \ref{fig:nonlin_ic_fourier} show that small-scale density perturbations grow only when the isochoric mode is linearly stable and not when it is linearly unstable. In the latter case, only large scale isochoric perturbations grow, which achieve pressure equilibrium much later after a sound-crossing time across the mode, as the cold under-pressured region collapses.

\subsubsection{Power spectra}
We calculate the power spectra of the various runs at late times in Figure \ref{fig:nonlin_ic_fourier} and show it in the bottom panel. We use uniform binning in $\log (k)$. We calculate the power spectrum as
\begin{equation}
    E_\rho(k_i) = \sum_{k_i \leq k < k_{i+1}} \frac{|\widetilde\rho(k)|^2}{(k_{i+1}-k_i)}, 
\end{equation}
where $k_i$s are bin boundaries and $E_\rho(k_i)$ is the power in the $k_i$ bin. We note that all runs with unstable isochoric modes have a higher fraction of their power in low $k$s in comparison to the runs with stable isochoric modes. This indicates the presence of 
dense clouds separated by a larger distance for an unstable isochoric mode and small-scale multiphase gas with a stable isochoric mode. Note that the power spectra in the bottom row of Fig.~\ref{fig:nonlin_ic_fourier} lack any prominent peaks corresponding to length scales that can be seen in the density profiles. This is similar to the power spectrum of a square wave of varying width that also shows a broad power-law distribution.

\section{Discussion \& conclusions}
\label{sec:disc}
In this paper we 
study the linear and nonlinear evolution of the isobaric and isochoric local thermal instability in an optically thin plasma 
and its dependence on the gas temperature and metallicity. The role of metallicity on thermal instability has not been explored earlier.
The fragmentation of isochoric clouds and the broad implications of such clouds on the observations of warm/cold gas have been discussed in \citealt{McCourt2018} in detail. 
This paper invokes a characteristic length scale for the emergence of stable cloudlets ($l_{\rm cloud} \sim {\rm min} [c_{\rm s} t_{\rm cool}]$; see also \citealt{Burkert2000}) 
as a result of shattering. More recently, thermal instability on large scales has been studied by \citealt{Waters2019} and \citealt{Gronke2020}, who do not find much evidence for shattering. 
We 
revisit this problem using linear theory and high resolution 1D hydrodynamic simulations and find that the linear behavior also determines the nonlinear evolution (see Fig. \ref{fig:multiscale_diag}).  

While the cooling function ($\Lambda [T,Z]$) varies with both gas temperature and metallicity, we show that the temperature variation of the cooling function plays a more crucial role than the metallicity variation in the growth and nonlinear evolution of cold gas. Although we use high-resolution 1D hydrodynamic simulations, we note that 
the study of condensation due to thermal instability 
requires a very high resolution (APPENDIX \ref{app:numdiff}). 
Large volume cosmological/non-cosmological simulations still cannot resolve  
the relevant small 
scales. 
Even in our 1-D study we have to use a combination of small and large box simulations to study the appropriate length scales. 

\subsection{Metallicity \& thermal instability}

Using linear theory and nonlinear 1D simulations, we show that the thermal instability does not explicitly depend on the variation of the cooling function with metallicity ($\Lambda_Z=\partial \ln \Lambda (T,Z)/\partial \ln Z$). Of course, a higher metallicity medium has shorter cooling and thermal instability timescales because of the metallicity dependence in the cooling function (Eq. \ref{eq:CF}). Unlike metallicity, the variation of the cooling function with temperature ($\Lambda_T=\partial \ln \Lambda (T,Z)/\partial \ln T$) explicitly features in the expression for the isobaric thermal instability growth rate, and hence determines the multiphase condensation timescale. This difference of thermal instability with respect to $\Lambda_T$ and $\Lambda_Z$ can be understood intuitively. In isochoric conditions (when the sound-crossing time across a disturbance is longer than the cooling time), the overdense region cools but there is negligible background compression and hence no change in density. In absence of background fluid motion, the passive scalar metallicity of the cooling blob remains unchanged. The perturbed temperature evolve as $d\delta T/dt \propto -\Lambda_T \delta T $, which implies isochoric instability for $\Lambda_T<0$ (see Eq. \ref{eq:isochoric}). In isobaric conditions, the overdense region maintains pressure equilibrium with the background and there is convergence on to it. But in the local limit (fluctuation scale $\ll$ metallicity gradient scale), unlike temperature and density, the passive scalar metallicity of the cooling blob does not change. The temperature fluctuation satisfies $d\delta T/dt \propto -\delta (\Lambda/T^2) \propto -(\Lambda_T-2) \delta T$, resulting in a growth rate given by Eq. \ref{eq:isobaric}. Thus, there is no explicit dependence of the growth rate on $\Lambda_Z$ in either the isochoric or isobaric limits.

In past, variation in metallicity has been invoked to suppress emission of metal lines below a keV (\citealt{Morris2003}), as required by observations. But ours is the first attempt to  explicitly study the dependence of metallicity on the linear and nonlinear evolution of thermal instability. Nonlinearly, in the simulations with background metallicity gradient we find jumps in the metallicity across cold gas (bottom right panels of Figs. \ref{fig:nonlin_ib}-\ref{fig:nonlin_ic_outflow}). These occur because metallicity evolves as a passive scalar advected by the background flow. Because of this, the metallicity map reveals the regions in the diffuse gas from which the cold gas is condensing. The mean metallicity of the cold gas is the mass-weighted metallicity of the gas that condenses from the surrounding regions. Thus if the background hot gas metallicity has variation, the metallicity of the cold gas is closer to that of the diffuse region from which it is condensing (Fig. 7 in \citealt{Nelson2020} shows correlations between the cold gas density and the background metallicity). The cold gas is not always expected to occur at the local metallicity maxima but where the cooling time is short (e.g., condensation can occur in high density but low metallicity regions and not in high metallicity dilute gas).


\subsection{Absence of fragmentation cascade in isochoric clouds}
\label{sec:frag_isochoric}
\citet{McCourt2018} invoke a faster process to reach pressure equilibrium for large clouds which cannot communicate pressure forces rapidly because they are in the isochoric regime with the sound crossing time longer than the cooling time. Such a cloud is expected to break into a cascade of smaller isobaric fragments until these cloudlets reach a stable 
characteristic scale $\sim {\rm min}(c_{\rm s} t_{\rm cool})$ near $10^4$ K. We do not see a cascade of fragmentation in any of our nonlinear 1D simulations. While 1D simulations lack hydrodynamic phenomena such as Rayleigh-Taylor/Kelvin-Helmholtz instabilities and mixing, the above picture of hierarchical fragmentation due to cooling is expected to be captured in 1D, if present. Although we do not see the fragmentation of the dense isochoric regions, we do see the growth of isobaric perturbations at small scales in the nonlinear state if the isochoric mode at the background temperature/metallicity is linearly thermally stable. So in this case, we see cold gas arising at small scales but this is different from what is envisaged in \citet{McCourt2018}.

\citet{Gronke2020} carry out 3D simulations of isochoric clouds collapsing due to radiative cooling. Even in their simulations the cold clouds, in all regimes, cool and collapse nonlinearly without fragmenting in the process. However, fragmentation and shattering are seen {\it after} the collapse of cloud to a small volume as it attains the stable temperature.\footnote{as seen in the movies associated with the paper: \url{http://max.lyman-alpha.com/shattering/}. }
The late-time shattering is a result of the vorticity generated by the
Richtmyer-Meshkov instability produced due to the interaction of the rebounding shock and the density gradients in the multiphase cloud (\citealt{Richtmyer1960, Meshkov1969}). This instability causes a cascade of smaller cold gas structures in their 3D simulations. Hence, these small-scale clouds are also not produced by the mechanism of \citet{McCourt2018}.
\citet{Gronke2020} find small-scale structure in the cold gas for $\chi \gtrsim 300$ (the final temperature ratio of the background and the stable phase). 
At later times the small-scale cold clouds coagulate rather than fragment. In reality, the structure of cold gas is determined by the complex interplay of cooling and heating in a turbulent medium (\citealt{Banerjee2014,Mohapatra2019}), with both fragmentation and coagulation happening simultaneously.


\subsubsection{Nonlinear initial conditions}
\label{sec:nonlin_test}

\begin{figure}
    \includegraphics[width=\figsize\textwidth]{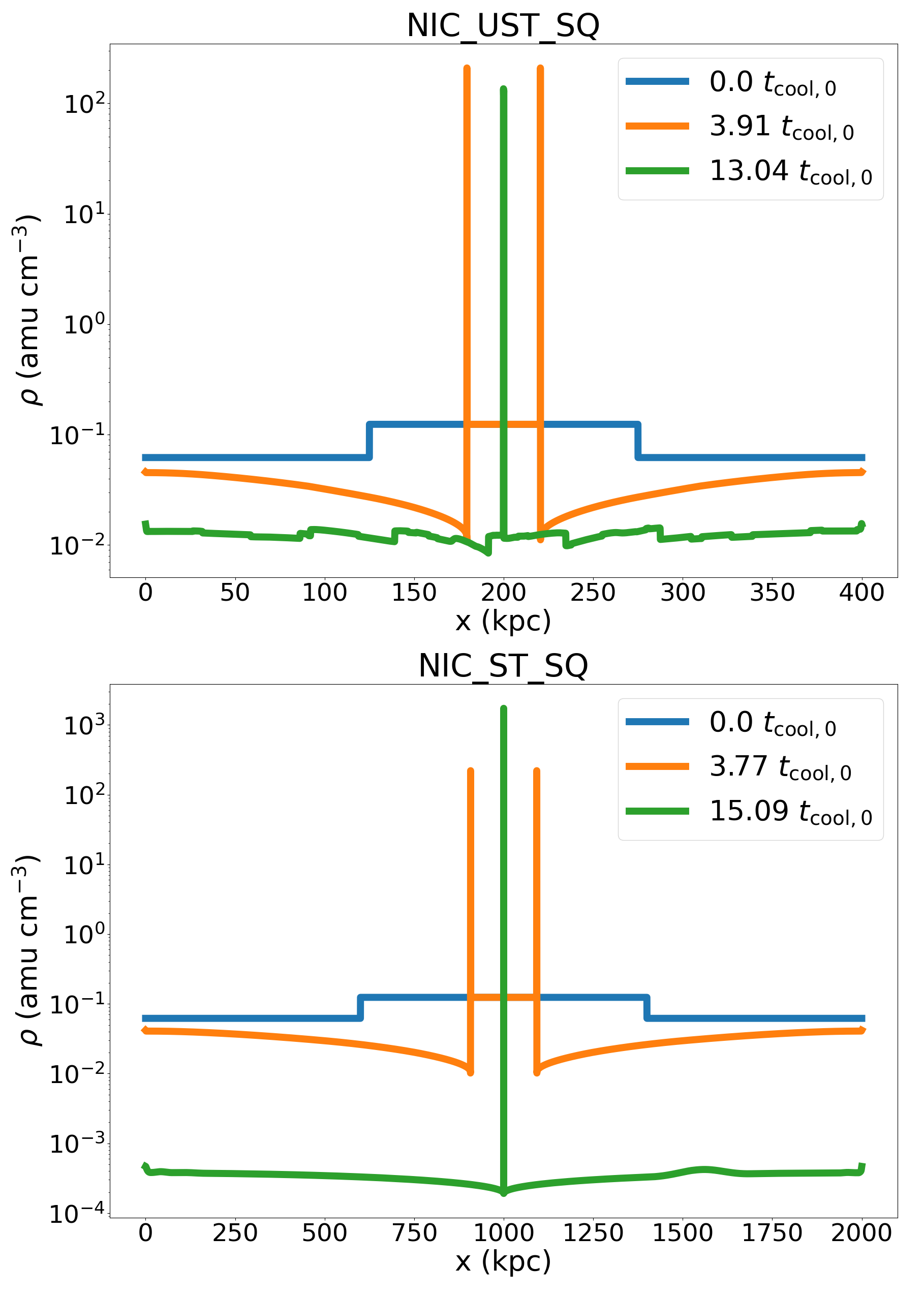}
    \centering
    \caption{Density profiles at different times for the runs initialized with large density square waves: NIC\_UST\_SQ (unstable isochoric regime; top) and NIC\_ST\_SQ (stable isochoric regime; bottom) at different times, given by different colors. The initial pressure is uniform across the box with overdensity temperatures set at $1\times10^7~\rm K$ (NIC\_UST\_SQ) and $3\times10^7~\rm K$ (NIC\_ST\_SQ), and the ambient temperatures at $2\times10^7~\rm K$ (NIC\_UST\_SQ) and $6\times10^7~\rm K$ (NIC\_ST\_SQ). We find that with nonlinear initial conditions, the evolution is similar regardless of stability of the isochoric mode.}
    \label{fig:nonlin_initial}
\end{figure}

\citet{Gronke2020} point out that there is no difference in the fragmentation threshold $\chi_f \sim 300$ for linear thermal instability and non-linearly growing overdensities. Following this, we test if our criterion of isochoric stability/instability depends on the initial amplitude of perturbations. To do so, we initialize a square wave density pulse with constant pressure across the domain and carry out stable and unstable isochoric simulations (ST and UST; in both cases, the setup is isochorically stable and unstable respectively for all temperatures between initial minimum and maximum temperatures). We find that, 
unlike when we start from linear perturbations, the dependence of nonlinear evolution on the linear stability of the isochoric mode breaks down if we start with non-linear perturbations (Fig.~\ref{fig:nonlin_initial}). Hence, the 
isochoric stability/instability is expected to be important only for clouds  growing from small density perturbations, $\delta \rho/\rho \lesssim 1$. This result is expected because with spatially constant heating the diffuse gas is heated overall rather than being in thermal balance. This 
precludes the applicability of our result on connection between linear behavior and nonlinear evolution for condensation around pre-existing dense gas like accreting filaments in gaseous halos and tails of jelly-fish galaxies. 
Our results are applicable for gas cooling out of a relatively uniform medium, e.g., the cool ICM core, small-amplitude perturbations in outflows. 

If we simply accept that \citet{Gronke2020}'s threshold 
applies for the origin of small-scale clouds (as discussed in section 3 of \citealt{Gronke2020}), and the cloud finally comes to pressure equilibrium with surroundings, then $\chi_f = 300$ translates to a background hot gas at T$\sim 10^7~{\rm K}$ (assuming cloud cools to the floor temperature $T_{\rm flr} \sim 4\times 10^4~{\rm K}$). The 
ICM has a similar or a larger virial temperature, so it is expected to be prone to multiphase fragmentation. Relating this to our small-amplitude simulations, it is interesting to note that the ICM is isochorically stable 
and is expected to show rapid growth of small-scale isobaric clumps. However, this may also be a coincidence since Fig. \ref{fig:nonlin_initial} shows a rather similar evolution for nonlinear initial perturbations, irrespective of the stability or instability of isochoric modes. The origin of $\chi_f \sim 300$ and its robustness need further investigation.

\subsection{Is there a characteristic length scale in cold gas?}

A key question in the nonlinear evolution of thermal instability, as it produces multiphase gas, is whether there is a characteristic scale for the cold gas. Figure \ref{fig:multiscale_diag} outlines the nonlinear evolution of the large isochoric clouds when the isochoric mode is unstable (the left channel) and stable (the right channel). In the former case that applies to the CGM with the virial temperature $\sim 10^6$ K (see Fig.\ref{fig:growth_rate_theory}), the large isochoric cold clouds collapse on a long sound crossing time ($\sim \lambda/ c_{s, \rm cold} \sim 1~{\rm Gyr} [\lambda/10~{\rm kpc}$] for the $10^4$ K cold phase). Such large-scale, pre-collapse cold clouds at almost the density of the hot ambient medium may explain the observation of the low density/pressure $\sim 10^4$ K gas in \citet{Werk2014} (see their Fig. 12).

The possibility of finding a characteristic length scale for cold gas is numerically demanding (see APPENDIX \ref{app:numdiff}). 
Cooling gas can shrink down to very small volumes as it attains pressure equilibrium, so that the resolution required to resolve the cold gas is enormous. A region of length $l$ in the hot phase would eventually collapse to $l\chi^{-1}$ in 1D, a factor of $\sim 1000$ for the fiducial ICM parameters. The cold gas scale is not so small $\sim l \chi^{-1/3}$ for isotropic collapse in 3D. Given this and other crucial differences in 1D and 3D, the question of characteristic length in the CGM can only be answered quantitatively in 3D. Here we perform high resolution 1D simulations to investigate the question of characteristic length of cold gas in the context of thermal instability in 1D.

\begin{figure*}
    \includegraphics[width=18cm]{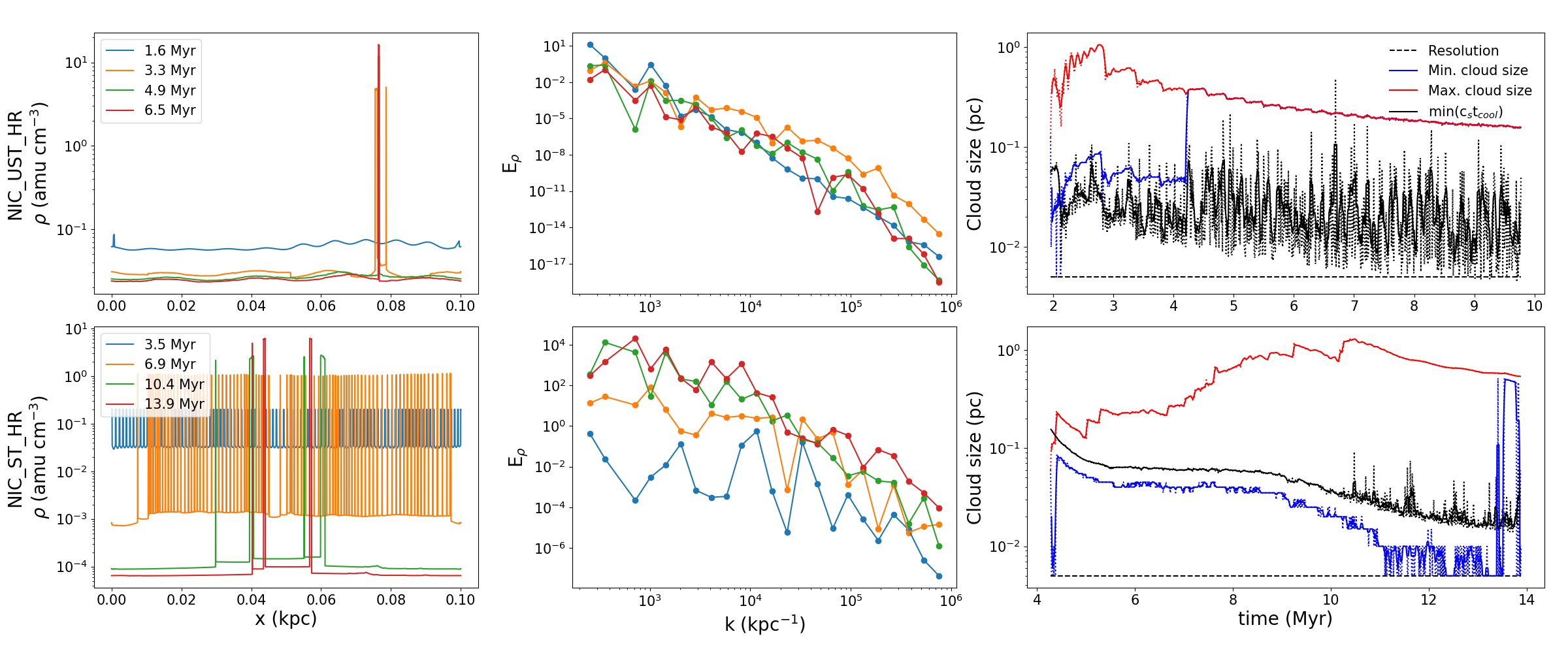}
    \centering
    \caption{The density profiles (left column) and the density power spectra (middle column) at different times for the isochoric unstable (top panels) and isochoric stable (bottom panels) low-temperature high-resolution (20000 grid points) simulations with isochoric+isobaric initial fluctuations. The right panels 
    show the size of the smallest and largest cold ($T<10^{4.1}$ K) clouds, the characteristic length scale of cold gas (min[$c_s t_{\rm cool}$]), and the grid size (dashed line);
    the solid lines are the rolling averages of the actual values (shown as dotted lines) over 5 time steps centered at each time step. Many more dense clouds separated by small distances (reflected in a shallower power spectrum), which eventually merge, are seen in the stable isochoric run. A single dense cloud is seen for the unstable isochoric run after $\approx 4$ Myr. In both cases, only a 
    few dense clouds exist at late times as a result of mergers. A reduction in the largest cloud size at late times is likely a result of the increase in the ambient pressure because of heating of the diffuse medium.
    }
    \label{fig:cloud_size_t}
\end{figure*}

We carry out two simulations, 
one with a background temperature $\approx 4 \times 10^{4}$ K for a stable isochoric cloud (see the stability trough in Fig. \ref{fig:cubic_intermediate} around such temperatures) and 
another at temperature $\approx 3 \times 10^{5}$ K for an unstable isochoric cloud. The two runs have the same intermediate wavenumber $k = 2\pi/100~{\rm pc}^{-1}$. Resolving the smallest length scales is difficult if we initialize the asymptotic ($kc_s \ll 1$) isochoric modes at a higher temperature since the initial cloud sizes are very large. 
We carry out simulations of the two isochoric cases in a $100~{\rm pc}$ box to follow the evolution of the largest and the smallest cloud sizes in the simulations (see the right column in Fig. \ref{fig:cloud_size_t}). These simulations have 20000 grid cells. The initial profiles in both cases have small-scale isobaric eigenmodes ($\lambda = 1$ pc) seeded within the large-scale isochoric mode (these are similar to the mixed-eigenmode simulations of section \ref{sec:mixed_eig}). We also dynamically follow the the expected characteristic length scale, the minimum $l_{\rm cloud, shatter} \sim c_{\rm s} t_{\rm cool}$ (see \citealt{Burkert2000}) over the domain, as the clouds gradually cool down to lower temperatures. We verify that the results of Fig. \ref{fig:cloud_size_t} are similar whether we use a lower-order RK2/linear or a higher-order RK3/parabolic method.

In our 
simulations, $l_{\rm cloud, shatter} \sim {\rm min} (c_{\rm s} t_{\rm cool})$ is initially larger than the smallest overdensities. 
In the unstable isochoric run (top panels of Fig. \ref{fig:cloud_size_t}), we see that the smallest cloud follows the evolution of the smallest characteristic length scale till $\sim 4$ Myr. The size of the largest cloud is larger than $l_{\rm cloud, shatter} \sim {\rm min} (c_{\rm s} t_{\rm cool})$ throughout the entire time. The size of the biggest cloud intermittently reduces to slightly smaller size (possibly due to the pressure wave oscillations at the interface of the cloud and the surrounding in-falling gas; see \citealt{Waters2019}) but never reaches $l_{\rm cloud, shatter}$ or the grid scale. Note that the smallest and the largest clouds merge after 4 Myr and the resultant cloud remains stable at $\sim 5-10$ times larger size than $l_{\rm cloud, shatter}$.

In our 1D stable isochoric simulations (shown in the lower panels of Fig. \ref{fig:cloud_size_t}), we find that the initial overdensities are smaller than $l_{\rm cloud, shatter}$. 
As condensation proceeds, $l_{\rm cloud, shatter}$ reduces, closely followed by the smallest cloud in the simulation. 
Although our stable isochoric evolution indeed shows small-scale growth and the emergence of a characteristic scale for the smallest cloud, 
the isochoric cloud does not 
fragment. Thus, our fragmentation is different in nature than the picture of \citealt{McCourt2018}. In our simulations, small-scale features grow within a large cloud if and only if the linear isobaric mode is growing and the isochoric mode is decaying.

Figure \ref{fig:cloud_size_t} shows that the density profiles and the power spectra of the two isochoric simulations agree with our previous results (see Fig. \ref{fig:nonlin_ic_fourier}). Namely, a larger number of small-scale peaks in the stable isochoric case 
co-exist for a long time. The stable isochoric simulation has higher power in large $k$'s at all times, which verifies that small-scale growth is generally not seen within a growing isochoric mode. The smallest cloud-size often follows the characteristic length-scale (min[$c_st_{\rm cool}$]) scale for the transient state in both (stable/unstable isochoric) cases as this is the only relevant length scale in the problem. Moreover, at late times the small clouds merge and lead to much larger clouds than this scale. On the other hand, larger clouds are not uncommon (red lines in the right column). Thus our 1D simulations indicate that a characteristic length scale of cold gas may not generically exist. The length scale of cold gas will depend on the total mass of cold gas and the number of clouds (which decreases with time as the clouds merge; see Fig. \ref{fig:multiscale_diag}) that are in pressure equilibrium with the ambient medium at late times.
The cold clouds are very likely supported by non-thermal components such as magnetic fields and cosmic rays at a scale much larger than this characteristic length scale (e.g., \citealt{Sharma2010,Nelson2020}).

\subsection{Conclusions}
Following are the main conclusions of our 1D linear and nonlinear study:
\begin{itemize}
    \item From local WKB analysis, we find that the growth rate of the local thermal instability does not have an explicit dependence on metallicity. The growth rate with a higher metallicity is larger because the cooling time is shorter for a higher metallicity gas. 
    We verify the theoretical growth rates of both isobaric and isochoric modes with simulations in the linear regime. 
    \item We find two regimes of nonlinear evolution 
    depending on the linear stability of the 
    isochoric mode: we observe growth of cold gas with small separation for the isochoric mode when it is linearly stable; for linearly unstable isochoric cloud we do not observe cold gas separated by small scales in the nonlinear state. Our nonlinear evolution, even in the stable isochoric regime, is somewhat different from the one envisaged by \citet{McCourt2018} as we do not observe the dense isochoric cloud break into several isobaric fragments. Instead, we observe the growth of isobaric modes at small scales.
    We see no condensing cold gas at small scales in runs with unstable isochoric modes, as suggested in \citealt{Waters2019}. 
     In the unstable isochoric runs, we see the formation of a large cloud that shrinks and eventually forms an isobaric overdense region.
     The isobaric mode is thermally unstable for all temperatures $\gtrsim 10^4$ K, but the isochoric mode is stable for a temperature $\gtrsim 10^7$ K and for small ``islands'' between $10^4$ to $10^7$ K (see Fig. \ref{fig:growth_rate_theory}). The production of small-scale cold gas for stable isochoric modes may explain the result from the 3D simulations of \citet{Gronke2020}, who find fragmentation of cold gas for density contrast $\chi \gtrsim 300$ (corresponding to a temperature of $\gtrsim 10^7$ K). However, we also find that the correspondence between nonlinear evolution and the linear stability of isochoric modes breaks down if the initial density perturbations are already nonlinear ($\delta \rho/\rho \gtrsim 1$).
     We test the robustness of our result by using different initial conditions such as mixed eigenmodes and random noise. We also compare the power spectra of density profile in different cases. We find the results in each case to be consistent with our pure eigenmode simulations. 
    \item The observed cold gas in our nonlinear simulations does not generically show a characteristic scale even when we resolve the expected characteristic cold gas radius of min$(c_s t_{\rm cool})$ (see Fig.\ref{fig:cloud_size_t}). The minimum cloud size in our 1D simulation in the growing/transient state is comparable to  min$(c_s t_{\rm cool})$ (as this is the only length-scale in the problem) but merger of cold clouds leading to larger clouds is common at late times. 
    \item The passive scalar nature of metallicity can help us connect the cold gas with its source in the diffuse CGM. The condensing gas inherits the metallicity of the diffuse gas from which it is condensing.
\end{itemize}

%
%
\section*{Acknowledgements}
HKD acknowledges support from a fellowship provided by the Kishore Vaigyanik Protsahan Yojana (KVPY) scheme of the Department of Science and Technology, Government of India. PPC acknowledges the European Research Council (ERC) for support under the European Union\textquotesingle s Horizon 2020 research and innovation programme (project DISKtoHALO, 
grant 834203). PS thanks David Buote and Francois Mernier for useful email exchanges that motivated the analysis of thermal instability with metallicity variations. PS also acknowledges a Swarnajayanti Fellowship from the Department of Science and Technology, India (DST/SJF/PSA-03/2016- 17). We thank Peng Oh, Tim Waters and an anonymous referee for useful suggestions.
%


\section{Data Availability}
All the relevant data associated with this article will be shared on reasonable request to the authors.



\bibliographystyle{mnras}
\bibliography{References} 

%
%
%
\appendix

\section{Numerical diffusion at high wavenumbers}
\label{app:numdiff}

\begin{figure}
    \includegraphics[width=\figsize\textwidth]{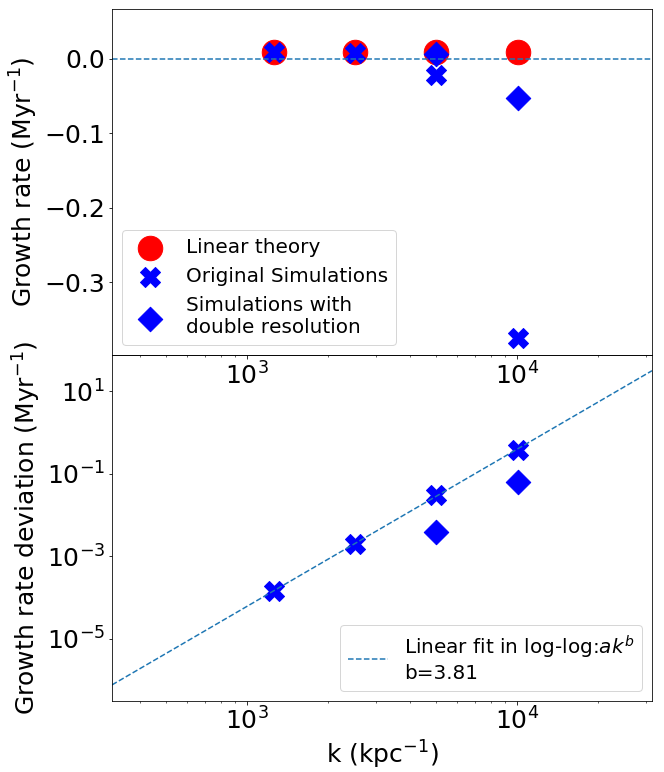}
    \centering
    \caption{{\em Top panel}: Linear growth rates and the rates measured from numerical simulations initialized with large-$k$ eigenmodes. {\em Bottom panel}: The deviation of the measured growth rate from the theoretical value for different $k$s. For a given resolution, the deviation from the theoretically expected growth rates increase with the wavenumber ($k$), which follows a power-law in $k$. For higher resolutions, the power remains the same but with a smaller coefficient.}
    \label{fig:vhighk_growth}
\end{figure}

In some of our simulations with a small growth rate of the local thermal instability, we observe large deviation between the theoretical and numerically measured growth rates. In all cases the growth rate in simulations is lower than the theoretically expected value. In some cases we measure damping of the eigenmode instead of growth, even though linear analysis suggests otherwise. 
For a given resolution, the deviation from the theoretical expectation increases with the wavenumber ($k$) as shown in the first panel of Fig.~\ref{fig:vhighk_growth}.

We find that the deviation from the theoretical growth rate follows a power-law in $k$, as shown in the second panel of Fig.~\ref{fig:vhighk_growth}. This deviation is due to numerical diffusion, which is expected to introduce a 
dissipation rate increasing rapidly with $k$ (a constant numerical diffusion should give a damping rate $\propto k^2$); the numerically measured deviation is steeper. On increasing resolution, the effects of numerical diffusion decrease, resulting in a better agreement between the theoretical and numerical values, as seen in  Fig.~\ref{fig:vhighk_growth}. Because of the long cooling time, compared to the sound-crossing time across a grid-cell ($\Delta x$ is the grid-size) in some cases, the numerical dissipation rate $\sim (c_s \Delta x) k^2 $ can significantly damp the growth rate and it is important to adequately resolve the high-$k$ modes and to conduct convergence studies.

\bsp	
\label{lastpage}
\end{document}